\documentclass[12pt,twoside]{article}
\usepackage{amssymb,amsmath,bbm}
\usepackage{latexsym}
\usepackage[english]{babel}
\usepackage{longtable}
\usepackage{epsfig}
\usepackage{graphicx,psfrag,subfigure}
\graphicspath{{images/}}

\numberwithin{equation}{section}
\title{{\large Lecture Notes:} {\normalsize Fundamental Problems in Statistical Mechanics XI,
\vspace*{-8pt}\\
Leuven, September 4 - 16, 2005}\bigskip\bigskip\bigskip\bigskip\bigskip\\
{\Large {\bf Exact solutions for KPZ-type growth processes,\\ random
matrices,\\ \vspace*{-3pt} and equilibrium shapes of crystals} \bigskip \bigskip}}
\author{{\large Herbert Spohn} \medskip\\ {\normalsize Physik Department and Zentrum Mathematik,}\\
 {\normalsize Technische Universit\"{a}t M\"{u}nchen, Boltzmannstr. 3, D-85747 Garching} \\{\normalsize e-mail: {\tt spohn@ma.tum.de}}}
\date{{\normalsize December 1, 2005}}

\begin{document}
\maketitle
\vspace*{6cm}
\noindent
{\bf Abstract}: Three models from statistical physics can be analyzed by employing space-time determinantal processes:
(1)  crystal facets, in particular the statistical properties of the facet
  edge, and equivalently tilings of the plane,
(2) one-dimensional growth processes in the
  Kardar-Parisi-Zhang universality class and directed last passage
  percolation,
 (3) random matrices, multi-matrix models, and Dyson's Brownian motion. We explain the method and survey results of physical interest.
\newpage

\section{Introduction}\label{sec.1}
\setcounter{equation}{0}

Exactly solvable models from Equilibrium Statistical Mechanics in
two  space-dimen\-sions are a continuing source of fascinating
research. Historically the Onsager solution of the two-dimensional
Ising model stands out. But the field has progressed. The most
recent advance is conformal field theory and its probabilistic
underpinning through the link to the Schramm-Loewner evolution
(SLE) \cite{Ca}.

Going back to the Ph.D.~thesis of Ising, one conventional starting
point for an exact solution in equilibrium statistical mechanics
is to write the Boltzmann weight as a product of matrices. Each
factor is referred to as \textit{transfer matrix}. In case of a
spin system on the two-dimensional lattice $\mathbb{Z}^2$ the
matrix elements of the transfer matrix are labeled by spin
configurations in adjacent columns. Under favorable circumstances
the transfer matrix can be rewritten as the configuration space
kernel of $\exp[-H]$, where $H$ is bilinear in fermionic
creation/annihilation operators labeled by the sites of a single
column. In such case one says that the model can be solved through
a mapping to \textit{free fermions}. The problem of finding the
largest eigenvalue of the $2^N\times 2^N$ transfer matrix is
reduced to the eigenvalue problem of an $N\times N$ matrix. For
the 2D Ising model the free fermion method was discovered by Lieb,
Mattis and Schultz \cite{LMS}. Under less favorable circumstances
one can still extract information from the transfer matrix by more
sophisticated methods, like the Bethe ansatz, the Yang-Baxter
equations, and the technique of commuting transfer matrices. We
refer to \cite{LW,Ba} for details. For the purpose of our lectures
the free fermion method with suitable extensions will do.

To be more concrete, and to anticipate some features to be
developed in much greater depth further on, let us explain the
free fermion method in the context of the ANNNI model \cite{Vi},
to say the 2D anisotropic next nearest neighbor Ising model. The
spins, with values $\pm 1$, are located at the sites of the square
lattice $\mathbb{Z}^2$. Because of anisotropy the properties in
the 1- and 2-direction are very different. It is then convenient
to consider the 1-axis as fictitious ``time'' and the 2-axis as
``space''. Along the time direction the spins interact via the
ferromagnetic nearest neighbor coupling $J_0$, $J_0>0$, while in
the space direction there is a nearest neighbor ferromagnetic
coupling $J_1$ and a next nearest neighbor antiferromagnetic
coupling $J_2$. We set $J_2=-J_1/2<0$. The ground state is then
highly degenerate. It consists of alternating strips of $+$ spins
and $-$ spins parallel to the time direction with a width of 2 or
larger.

\begin{figure}[t!]
  \begin{center}\hfill
  \subfigure[]{
  \includegraphics[height=5cm]{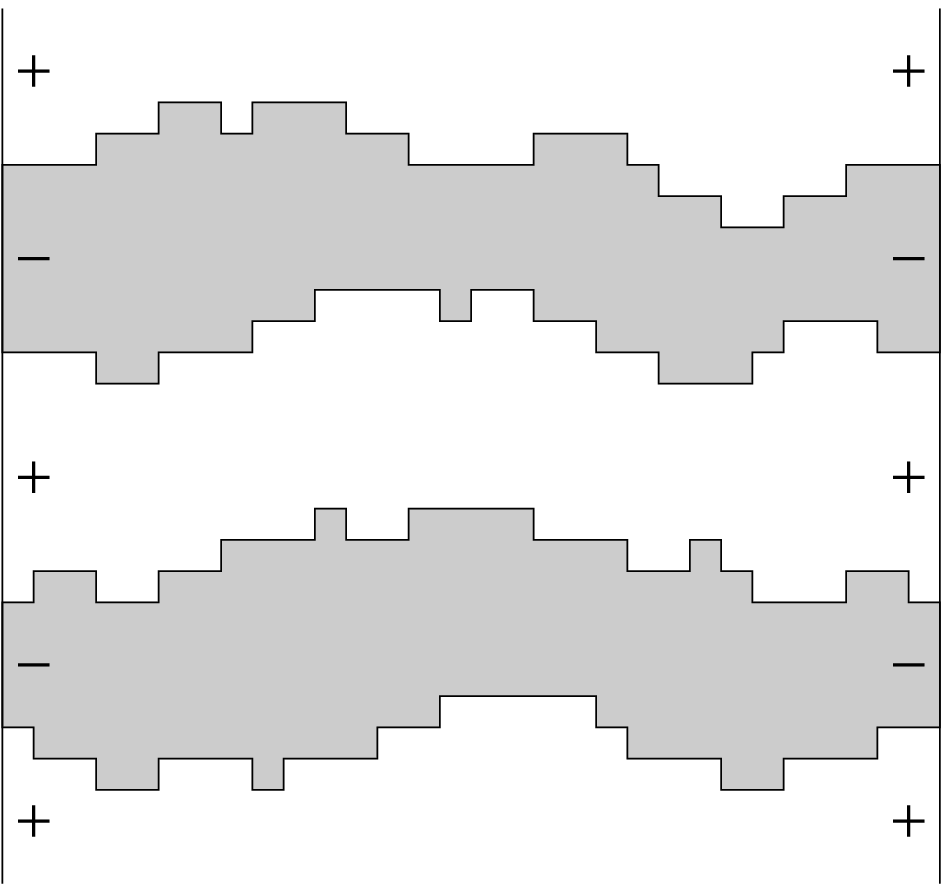}}\hfill
  \subfigure[]{
  \includegraphics[height=5cm]{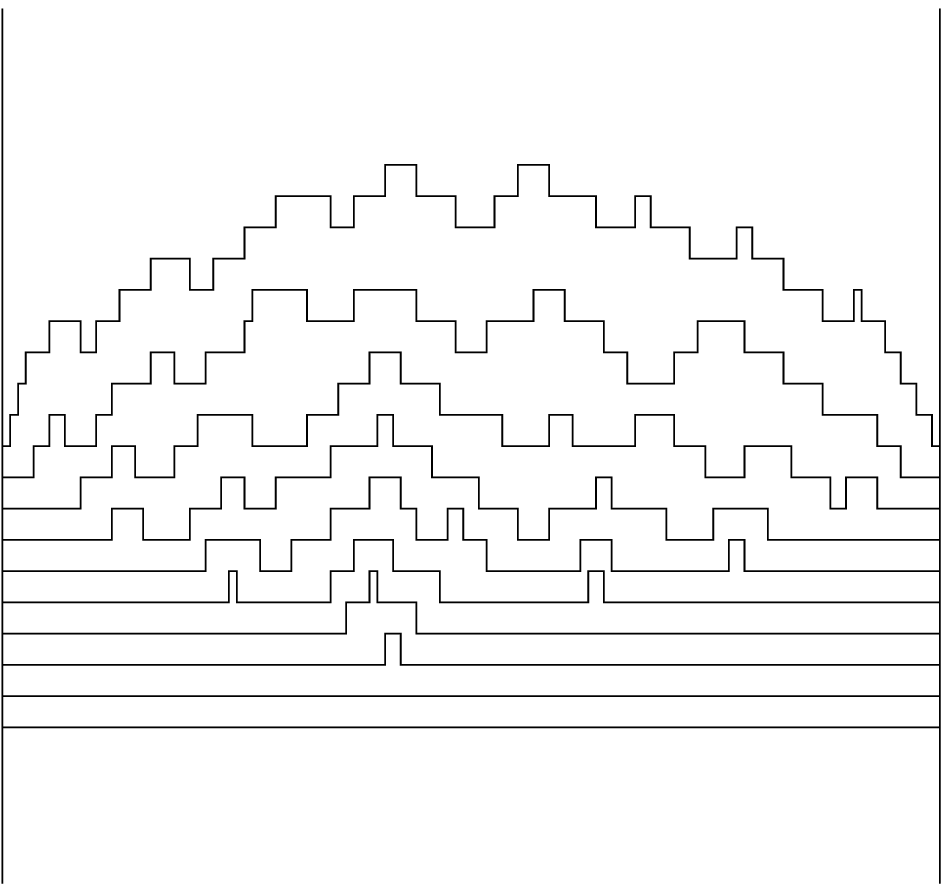}} \hfill \phantom{}
  \caption{(a) Domain walls of the ANNNI model and (b)
line ensemble of the PNG growth model. Note that for the line
ensemble (b) there is a top line. Mostly, our focus will be on the
statistical properties of the top line.}\label{Fig1}
\end{center}
\end{figure}

As the temperature is increased, the domain walls, i.e., the lines
separating the $+$ and $-$ domains, become thermally rough. In
approximation we postulate that a domain wall has up-steps and
down-steps of size 1 only and has no overhangs, see Figure
\ref{Fig1}(a). To have easier comparison with the models to be
discussed below, the distance between neighboring domain walls  is
diminished by one lattice spacing. Then we obtain the following
\textit{line ensemble}: There are $M$ domain walls $t\mapsto
x_j(t)$, $j=1,\ldots, M$, with $x_j\in[-N,\ldots,N]$,
$t\in[-N,\ldots,N]$. Each domain wall has steps at most of size 1,
$x_j(t+1)-x_j(t)=0,\pm 1$ for all $j,t$, and the domain walls
satisfy the \textit{non-crossing (non-intersecting) constraint}
$x_j(t)<x_{j+1}(t)$. To an admissible configuration of domain
walls one assigns a Boltzmann weight. A single step of the domain
wall has weight $e^{-J_0/k_\mathrm{B}T}$. Thus the Boltzmann
weight for a configuration of admissible domain walls is
\begin{equation}\nonumber
\exp{\big[}-\tfrac{J_0}{k_\mathrm{B}T}{\textrm{(number of
up-steps and down-steps)}}{\big]}\,.
\end{equation}

So far we have imposed a hard wall boundary condition in the space
and free boundary conditions in the time direction. One could also
require periodic boundary conditions in the $t$-direction. A
further popular choice are in addition chiral boundary conditions
along the space direction. Then the domain walls have a nonzero
slope on average.

The low temperature phase diagram of the ANNNI model, in the
approximation just explained, can by analyzed using free fermions.
Since we will have ample opportunity to explain the method, no
details are needed now. Let me emphasize that, while anisotropic,
the ANNNI model as defined is translation invariant. Thus the
interest is in the free energy per unit volume and in the
two-point function which depends only on the relative distance of
the two points. In contrast, the topic of my lectures is concerned
with line ensembles which are \textit{inhomogeneous} both in space
and time. Figure \ref{Fig1} illustrates the difference.
Over the past six years, for some instants even much further
back, it has been recognized that there are three physically very
distinct systems which can be analyzed through
\textit{inhomogeneous line ensembles}, namely
\begin{itemize}
  \item crystal facets, in particular the statistical properties of the facet
  edge, and equivalently tilings of the plane,
  \item one-dimensional growth processes in the
  Kardar-Parisi-Zhang universality class and directed last passage
  percolation,
  \item random matrices, multi-matrix models, and Dyson's Brownian motion.
\end{itemize}
In the first item the description through a line ensemble is
rather immediate, as it is in the last item when focusing on
Dyson's Brownian motion. For growth processes the link turns out
to be more hidden.

Our plan is to explain,  in fair detail, for the three items
listed how the physical model is mapped to a line ensemble with
determinantal correlations. I want to provide the reader with a
feeling for the method, in particular its flexibility and its
limitations. Thereby one has not yet gained any understanding of
the statistical properties of the model, but one has reached the
``trail head'' for a hike towards the summit by means of rather
formidable asymptotic analysis. The details of the hike are well
documented in the literature and I will make no attempt to
duplicate. However, I will discuss of what one learns about
crystal facets and growth models beyond the specific ``exactly
solved'' model.

The main technical tool will be determinantal point processes, which are discussed {\it per se}
in the following section. The application to statistical mechanics models will occupy
the remainder of the survey. \medskip\\
{\bf{Acknowledgements.}} My article is based on the Ph.D. thesis
of Michael Pr\"{a}hofer \cite{Pr03} and on the Ph.D. thesis of
Patrik Ferrari \cite{Fth}. I am very grateful for their sharing of
insights and the ongoing collaboration. In addition, I thank
Patrik Ferrari for supplying the figures.


\section{Determinantal point processes}\label{sec.2}
\setcounter{equation}{0}

The correlation functions of a determinantal point process are
computable from a single correlation (two-point) kernel. In this
respect, although otherwise very different, determinantal point
processes are similar to Gaussian random fields. Therefore we
first recall briefly\smallskip\\
\textit{Gaussian processes}. In the discrete setting one
starts from the family of mean zero Gaussian random variables
$X_1,\ldots,X_N$. Their covariance matrix is
\begin{equation}\label{2.1}
C(i,j)=\langle X_iX_j\rangle\,.
\end{equation}
Clearly $C(i,j)=C(j,i)$ and $C\geq 0$ as a matrix. Conversely,
every such matrix is the covariance of a Gaussian process. Higher
moments are computable through the pairing rule
\begin{equation}\label{2.2}
\langle \prod^{2m}_{\ell=1}X_{j_\ell}\rangle=
\sum_{{\mathrm{pairings}}\,\pi(k),\pi'(k)}
\prod^m_{k=1}C\big(\pi(k),\pi'(k)\big)\,,
\end{equation}
where the sum is over all possible pairings of the indices
$j_1,\ldots,j_{2m}$.
 An index  is allowed to appear several times in the list.

Since also space-time processes will be in demand, we add the time
index $t\in[0,T]$. We have then the family of mean zero Gaussian
processes $X_1(t),\ldots,X_N(t)$. As before their covariance
matrix is
\begin{equation}\label{2.3}
C(i,s;j,t)=\langle X_i(s)X_j(t)\rangle
\end{equation}
considered as a positive operator on $\mathbb{C}^N\otimes
L^2([0,T])$. Except for being continuous, the time-index $t$ is on
the same footing as the space-index $j$. In particular, the
pairing rule (\ref{2.2}) is still valid.\smallskip

With this background let us discuss determinantal point processes
which is done in two steps, static and dynamic.\smallskip\\
\textit{Determinantal point processes, spatial part}. For
simplicity let us first assume space to be discrete. At each site
$j\in[1,\ldots,N]$ there is an occupation variable $\eta_j=0,1$,
where 0 stands for empty and 1 for occupied, which is the reason
for the name ``point process''. We prescribe a hermitian matrix
$R(i,j)$ such that $0\leq R\leq 1$. Then the joint distribution of
the $\eta$'s is defined through the moments
\begin{equation}\label{2.4}
\langle \prod^m_{k=1}\eta_{j_k}\rangle=\det
\big(R(j_\ell,j_{\ell'})\big)_{1\leq \ell,\ell'\leq m}\,,
\end{equation}
provided the collection of indices $\{j_1,\ldots,j_m\}$ has no
double points. As for Gaussians there is the single correlation
kernel $R$ which fixes the full probability distribution. However,
note that two different $R$ kernels may give rise to the same
$\eta$-distribution. In the applications below, we will meet a
real $R$ matrix which is self-similar to a symmetric matrix
$\widetilde{R}$, i.e., $R(i,j)=g(i)\widetilde{R}(i,j)g(j)^{-1}$.
Clearly, according to (\ref{2.4}), the moments do not depend on
whether they are computed from $R$ or $\widetilde{R}$.

To understand the connection to free fermions let us introduce the
fermion algebra $a_j$, $j=1,\ldots,N$, satisfying the
anticommutation relations
\begin{equation}\label{2.5}
\{a_i,a_j\}=0\,,\;\{a_i^\ast,a_j^\ast\}=
0\,,\;\{a_i,a_j^\ast\}=\delta_{ij}
\end{equation}
with the anticommutator $\{A,B\}=AB+BA$. Let
\begin{equation}\label{2.6}
H= \sum^N_{i,j=1}a^\ast_i \mathfrak{h}_{ij}a_j
\end{equation}
be a quadratic fermion operator with $\mathfrak{h}$ the $N\times
N$ one-particle Hamiltonian. $H=H^\ast$ is equivalent to
$\mathfrak{h}$ being hermitian, $\mathfrak{h}=\mathfrak{h}^\ast$.
We then set $\eta_j=a^\ast_j a_j$, the fermionic occupation
variables, and
\begin{equation}\label{2.7}
\langle \prod^m_{k=1}\eta_{j_k}\rangle= Z^{-1}
\mathrm{tr}\big[e^{-H}\prod^m_{k=1}a^\ast_{j_k}a_{j_k}\big]\,,\quad
Z= \mathrm{tr}\big[e^{-H}\big]\,.
\end{equation}
It is an exercise in anticommutators to verify that the moments of
the occupation variables are determinantal with correlation kernel
\begin{equation}\label{2.8}
R(i,j)=Z^{-1}\mathrm{tr}\big[e^{-H}a^\ast_i a_j\big]=
\big((1+e^\mathfrak{h})^{-1}\big)_{ji}\,.
\end{equation}

Later on a prominent quantity will be the probability for the set
$B\subset[1,\ldots,N]$ to be empty. To compute it one could use
(\ref{2.4}); however the fermions offer a shortcut. Let $\chi_B$
be the indicator function for the set $B$ and let $N_B= \sum_{j\in
B}a^\ast_j a_j$. Then, with $\mathbb{P}(\cdot)$ our generic symbol
for probability,
\begin{eqnarray}\label{2.9}
&&\hspace{-51pt}\mathbb{P}(\eta_j=0,j\in B)=
\lim_{\lambda\to\infty}Z^{-1}\mathrm{tr}\big[e^{-H}e^{-\lambda N_B}\big]\nonumber\\
&&\hspace{36pt}=\lim_{\lambda\to
\infty}(\det(1+e^\mathfrak{-h}))^{-1}
\det(1+e^\mathfrak{-h}e^{-\lambda\chi_B})\nonumber\\
&&\hspace{36pt}=\det(1-\chi_B R\chi_B)\,.
\end{eqnarray}

\begin{figure}[t!]
\begin{center}
\psfrag{0}[r]{$0$}
\psfrag{T}[l]{$T$}
\psfrag{t}{$t$}
\psfrag{x}{$x$}
\includegraphics[height=5cm]{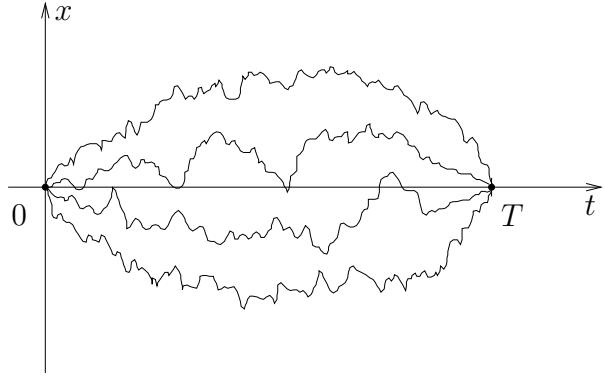}
\caption{Pinned Brownian motion conditioned not to intersect, also
referred to as watermelon ensemble.}\label{Fig2}
\end{center}
\end{figure}

To define determinantal processes over a continuum rather than
discrete space is as simple as for Gaussian processes \cite{S00}. Let
$\eta(x)$, $0\leq x\leq \ell$, be the point process. Every
realization is of the form
\begin{equation}\label{2.10}
\eta(x)=\sum^n_{j=1} \delta(x-y_j)\,,\quad y_j \in[0,\ell]\,.
\end{equation}
Here $n$ is arbitrary and $n=0$ means no point in $[0,\ell]$.
Furthermore we prescribe the correlation kernel through a linear
operator $R$ on $L^2([0,\ell])$ with $R=R^\ast$, $0\leq R\leq 1$,
and $\mathrm{tr}R<\infty$. Then, provided $\{x_1,\ldots,x_m\}$ has
no double points, one defines
\begin{equation}\label{2.11}
\langle \prod^m_{k=1}\eta(x_k)\rangle =\det
\big(R(x_\ell,x_{\ell'})\big)_{1\leq\ell,\ell'\leq m}\,.\smallskip
\end{equation}
\textit{Extended determinantal point processes}. To extend
determinantal processes to space-time, it is tempting to follow
the Gaussian example by introducing a space index $j$ and a time
index $t\in[0,T]$. Then, according to the definitions above,
$\eta_j(t)$ would be concentrated on a collection of points in
$[1,\ldots,N]\times[0,T]$, which is \textit{not} a line ensemble
of the form anticipated in the Introduction. Thus, to properly
guess the correct structure, we turn to the example of $n$
non-intersecting Brownian paths pinned at both ends to the origin,
which goes back to Karlin and McGregor \cite{KMG}. With such a
constraint and boundary condition typical paths resemble vaguely a
watermelon, compare with Figure \ref{Fig2}.
More precisely, for $j=1,\ldots,n$, $t\mapsto x_j(t)$, $0\leq
t\leq T$, is a one-dimensional Brownian motion pinned such that
$x_j(0)=0$, $x_j(T)=0$. In addition we impose the
\textit{constraint of non-crossing} as
\begin{equation}\label{2.12}
x_j(t)<x_{j+1}(t)\,,\quad 0<t<T\,,\quad j=1,\ldots,n-1\,.
\end{equation}
The corresponding random field over $\mathbb{R}\times[0,T]$ is
then
\begin{equation}\label{2.13}
\eta(x,t)=\sum^n_{j=1}\delta\big(x-x_j(t)\big)\,.
\end{equation}
If $T$ and $n$ are of the same order, then the constraint pushes
the top line a distance $T$ away from the origin, which is to be
compared with the typical $\sqrt{T}$ fluctuations for a single
Brownian bridge. Since the repulsion originates in a mere
constraint in the number of configurations, it is known as
entropic repulsion.

Let us first consider a single line, $n=1$. If the Gaussian
transition probability, from $x$ to $y$ in time $t$, is denoted by
\begin{equation}\label{2.14}
p_t(x,y)=\frac{1}{\sqrt{2\pi t}}e^{-(y-x)^2/2t}\,,
\end{equation}
one obtains
\begin{equation}\label{2.15}
\mathbb{P}_{0,0}\big(x_1(t)\in [x,x+d
x]\big)=p_T(0,0)^{-1}p_t(0,x)p_{T-t}(x,0)dx\,,
\end{equation}
where the subscript 0,0 reminds of the pinning to zero at both
ends. $p_T(0,0)^{-1}$ is the proper normalization.

Next let us consider two lines, but for simplicity only the point
statistics at fixed $t$. By the reflection principle, for
arbitrary starting points $x_j(0)=x_j$, $j=1,2$, $x_1<x_2$, 
and unconstrained end points, one
has the conditional probability
\begin{eqnarray}\label{2.16}
&&\hspace{-23pt}\mathbb{P}\big(x_1(t)\in [y_1,y_1+ dy_1],x_2(t)\in[y_2,y_2+ dy_2]\,
|\,x_1(0)=x_1,x_2(0)=x_2,\nonumber\\
&&\hspace{180pt}x_1(s)<x_2(s)\; \mathrm{for}\; 0\leq s\leq t\big)\\
&&\hspace{-13pt}=Z(t,x_1,x_2)^{-1}
\big(p_t(x_1,y_1)p_t(x_2,y_2)-p_t(x_1,y_2)
p_t(x_2,y_1)\big)\theta(y_2-y_1)dy_1dy_2\nonumber
\end{eqnarray}
with $\theta(u)=1$ for $u\geq 0$, $\theta(u)=0$ for $u<0$, and $Z$
the normalization. Therefore, with pinning at 0 and normalizing
partition function $Z(\delta)$,
\begin{eqnarray}\label{2.17}
&&\hspace{-26pt}\mathbb{P}_{0,0}\big(x_1(t)\in [x_1, x_1+dx_1],x_2(t)\in [x_2,x_2+
dx_2]\big)\nonumber\\
&&=\lim_{\delta\to
0}Z(\delta)^{-1}\big[p_t(0,x_1)p_t(\delta,x_2)-p_t(0,x_2)p_t(\delta,x_1)\big]
\nonumber\\
&&\hspace{36pt}\times\big[p_{T-t}(x_1,0)p_{T-t}(x_2,\delta)-p_{T-t}(x_2,0)
p_{T-t}(x_1,\delta)\big]dx_1dx_2\nonumber\\
&&\hspace{0pt}=\big(R(x_1,x_1)R(x_2,x_2)-R(x_1,x_2)R(x_2,x_1)\big)dx_1dx_2\,,
\end{eqnarray}
where
\begin{eqnarray}\label{2.18}
&&\hspace{-51pt}R(x,y)=p_t(0,x)p_{T-t}(y,0)
\big(p_T(0,0)^{-1}+Z^{-1}_2 xy\big)\,,\nonumber\\
&&\hspace{-51pt}Z_2=\int p_t(0,x)x^2p_{T-t}(x,0) dx\,.
\end{eqnarray}
Clearly, for fixed $t$ and $n=2$, the point process $\eta(x,t)$ of
(\ref{2.13}) is determinantal.

To extend to general $n$, still at fixed time $t$, it is
convenient to first introduce the Fermi field over $\mathbb{R}$
with creation/annihilation operators satisfying the
anticommutation relations
\begin{equation}\label{2.19}
\{a(x),a(x')\}=0\,,\;\{a^\ast(x),a^\ast(x')\}=0\,,\;
\{a(x),a^\ast(x')\}=\delta(x-x')\,,\;x,x'\in\mathbb{R}\,.
\end{equation}
To have a concise notation one also introduces the
harmonic oscillator Hamiltonian with frequency $1/t$ as
\begin{equation}\label{2.19a}
\mathfrak{h}_t=\frac{1}{2}\big(\!-\frac{d^2}{dx^2}+\frac{1}{t^2}x^2\big)\,,
\end{equation}
and its second quantization
\begin{equation}\label{2.20}
H_t=\frac{1}{2}\int a^\ast(x)
\big(\!-\frac{d^2}{dx^2}+\frac{1}{t^2}x^2\big)a(x)dx\,.
\end{equation}
We now proceed as in (\ref{2.17}), but for $n$ lines. In the limit
$\delta\to 0$ the left factor becomes $\psi^{(n)}_t$ and the right
factor $\psi^{(n)}_{T-t}$, where $\psi^{(n)}_t$ is the ground
state for $H_t$ with $n$ fermions. Therefore
\begin{equation}\label{2.20a}
\langle \prod^m_{k=1}\eta(x_k,t)\rangle= Z_T^{-1} \langle
\psi^{(n)}_{T-t}|\prod^m_{k=1}a^\ast(x_k)a(x_k)|\psi^{(n)}_t
\rangle_\mathcal{F}
\end{equation}
with $Z_T=\langle
\psi^{(n)}_{T-t}|\psi^{(n)}_t\rangle_\mathcal{F}$ and
$\langle\cdot|\cdot\rangle_\mathcal{F}$ the inner product in
fermionic Fock space. It is an exercise in anticommutators to
confirm that the right hand side of (\ref{2.20a}) is determinantal
with correlation kernel
\begin{equation}\label{2.21}
R_t(x,x')=  Z_T^{-1}\langle \psi^{(n)}_{T-t}|
a^\ast(x)a(x')|\psi^{(n)}_t\rangle_\mathcal{F}\,.
\end{equation}
One can use the normalized eigenfunctions $\varphi^t_j$ of
$\mathfrak{h}_t$, i.e., the Hermite functions, to express $R_t$ as
\begin{equation}\label{2.22}
R_t(x,x')=\sum^{n-1}_{j=0}\varphi^{\tilde{t}}_j(x)
\varphi^{\tilde{t}}_j(x')
\end{equation}
with $\tilde{t} = t(T-t)/T$. The correlation kernel is the
Hermite kernel of order $n$.

With these preparations, we are in a position to attempt our goal,
namely to extend to the point statistics at several times, e.g.,
to the joint probability distribution of $\eta(y_1,t_1)$ and
$\eta(y_2,t_2)$, $0<t_1<t_2<T$. Of course, following the examples
above, one can work out concrete formulas. They tend to be lengthy
and it is more transparent to emphasize the general principle. The
$n$-dependence turns out to be simple and we keep $n$ arbitrary,
but the reader is invited to verify for $n=2$. From the limit
$\delta\to 0$ at $t=0$ one obtains the ground state
$\psi^{(n)}_{t_1}$ at time $t_1$, and correspondingly from the
limit $\delta\to 0$ at $T$ one obtains $\psi^{(n)}_{T-t_2}$ at
time $T-t_2$. For the propagation from $t_1$ to $t_2$,
$\tau=t_2-t_1$, one has to use (\ref{2.16}) for general $n$, which
can be written as the position space kernel of $e^{-\tau G}$
restricted to the $n$-particle subspace, where
\begin{equation}\label{2.23}
G=-\frac{1}{2}\int a^\ast(x)\frac{d^2}{dx^2}a(x)dx\,.
\end{equation}
More explicitly, in position space,
\begin{equation}\label{2.23a}
\langle x_1,\ldots,x_n\,|\,e^{-\tau G}\,|\,y_1,\ldots,y_n\rangle
=\det\big(p_\tau(x_j-y_{j'})\big)_{1\leq j,j'\leq n}\,.
\end{equation}
Note that the free particle Hamiltonian $G$ provides the internal
propagation, while the harmonic oscillator Hamiltonian $H_t$
reflects the pinning. Therefore
\begin{equation}\label{2.24}
\langle\eta(y_1,t_1)\eta(y_2,t_2)\rangle =  Z_T^{-1}\langle
\psi^{(n)}_{T-t_2}|a^\ast(y_2)a(y_2) e^{-\tau G}a^\ast(y_1)a(y_1)|
\psi^{(n)}_{t_1}\rangle_\mathcal{F}\
\end{equation}
with $Z_T=\langle
\psi^{(n)}_{T-t_2}|\psi^{(n)}_{t_2}\rangle_\mathcal{F}$.

It is yet another exercise in anticommutators to verify that the
expression (\ref{2.24}) is determinantal with the correlation
kernel
\begin{equation}\label{2.25}
R(x,t;x',t')=
  \begin{cases}
  Z_T^{-1}\langle \psi^{(n)}_{T-t}|a^\ast(x)e^{-|t-t'|G}
 a(x')|\psi^{(n)}_{t'}\rangle_\mathcal{F} & \text{for}\; t\geq t'\,, \\
    - Z_T^{-1}\langle \psi^{(n)}_{T-t'}|a(x')e^{-|t-t'|G}
 a^\ast(x)|\psi^{(n)}_{t}\rangle_\mathcal{F} & \text{for}\; t< t'\,.
  \end{cases}
\end{equation}
(\ref{2.24}) generalizes to $m$ pairwise disjoint space-time
points $(y_1,t_1),\ldots, (y_m,t_m)$ which are ordered in time,
i.e., $0<t_1\leq t_2 \ldots\leq t_m<T$, as
\begin{equation}\label{2.26}
\langle \prod^m_{j=1}\eta(y_j,t_j)\rangle= \det
\big(R(y_j,t_j;y_{j'},t_{j'})\big)_{1\leq j,j'\leq m}\,.
\end{equation}
The moments are still of determinantal form. In contrast to the
static rule, for space-time points the time order must be
respected. On top the extended correlation kernel is \textit{not}
symmetric.

The general structure can be grasped even more clearly by
returning to the discrete space setting from (\ref{2.5}) above. In
addition to the static Hamiltonian (\ref{2.6}), there is the
generator
\begin{equation}\label{2.27}
G=\sum^N_{i,j=1}a^\ast_i \mathfrak{g}_{ij}a_j
\end{equation}
for the time propagation. We define
\begin{equation}\label{2.28}
a_j(t)= e^{tG}a_j e^{-tG}\,,\; a^\ast_j
(t)=e^{tG}a^\ast_je^{-tG}\,.
\end{equation}
Then the dynamic extension of the static correlation kernel $R$ from
(\ref{2.8}) is given through
\begin{equation}\label{2.29}
R(j,t;j',t')=
  \begin{cases}
 Z^{-1}\mathrm{tr} \big[e^{-H}e^{-TG}a^\ast_j(t)
 a_{j'}(t')\big] & \text{for}\; 0\leq t'\leq t\leq T\,, \\
 - Z^{-1}\mathrm{tr}\big[e^{-H}e^{-TG}a_{j'}(t')
 a^\ast_j(t)\big] & \text{for}\; 0\leq t< t'\leq T  \,,
  \end{cases}
\end{equation}
where
$Z=\mathrm{tr}\big[e^{-H}e^{-TG}\big]$.

In principle, the moments of the corresponding line ensemble are
still defined via (\ref{2.26}). For general $G$, one cannot expect
that the so defined moments come from a probability measure. In
fact, while $H$ can be arbitrary, except for $H=H^\ast$, the
conditions on $G$ are rather stringent, as will be explained now,
where we distinguish whether
space, resp.~time, is either continuous or discrete.\smallskip\\
(1) \textit{continuous time, continuous space}.
In this case $\mathfrak{g}$ must be the generator of a diffusion
process,
\begin{equation}\label{2.30}
\mathfrak{g}(t)=
-d(x,t)^2\frac{d^2}{dx^2}-a(x,t)\frac{d}{dx}+V(x,t)\,.
\end{equation}
As for the watermelon, the line ensemble is constructed from
independent lines with a weight determined by the propagator generated
through $\mathfrak{g}(t)$
and subsequently imposing the nonintersecting constraint. The
watermelon ensemble has constant diffusion and formally a strong
confining potential at $t=0$ and $t=T$, $V(x,t) =0$ otherwise. As will be explained,
the case $d=1$, $a=0$, $V(x)$ general corresponds to the eigenvalues of
hermitian multi-matrix models.
\smallskip\\
(2) \textit{continuous time, discrete space}. $\mathfrak{g}(t)$ is
the generator of a continuous time nearest neighbor random walk. If space is
$\mathbb{Z}$, then
\begin{equation}\label{2.31}
(\mathfrak{g}(t)\psi)_j=-r_+(j,t)\psi_{j+1}-r_-(j,t)\psi_{j-1} + (r_+(j,t)+r_-(j,t) +
V(j,t))\psi_j
\end{equation}
with rates $r_{\pm}\geq 0$. For the polynuclear growth model we will
encounter the simple random walk, for which
$r_+(j)=r_-(j)=\frac{1}{2}$ and $V(j)=0$. As before, the line
ensemble is obtained from independent lines by conditioning on
non-crossing.\medskip

For discrete time, there seems to be no complete classification.
Trivially, one can consider the cases (1) and (2) at discrete
times $t=n\tau$ only. In addition one finds one-sided exponential,
resp.~geometric jumps.\smallskip\\
(3) \textit{time discrete, space continuous}. Let us take the
ordered points $x_1<\ldots<x_n$. Then in an up-step the new
configuration $\{x'_j,j=1,\ldots,n\}$ has to satisfy $x_ j<x'_
j<x_{ j+1}$, $ j=1,\ldots,n$, formally $x_{n+1}=\infty$, and the
weight is $\prod^n_{ j=1} e^{-\delta(x'_ j-x_ j)}$, $\delta>0$.
Our example may look artificial, but does turn up in the analysis
of the totally asymmetric simple exclusion process.\smallskip\\
(4) \textit{time discrete, space discrete}. An obvious example are
nearest neighbor discrete random walks. To have a meaningful
non-crossing constraint odd and even sublattices must be properly
adjusted, see Example (i) below.  In case space is either
$\mathbb{Z}$ or $\mathbb{Z}_+$, the analogue of one-sided
exponential jumps are one-sided geometric jumps with weight $q^n$,
where $n$ is the jump size, $n\geq 0$, and $0<q<1$. One-sided
geometric jumps will show up for the Ising corner. In this model
time $t\in\mathbb{Z}$, for $t<0$ one has only up-steps, and for
$t\geq 0$ only down-steps, while the geometric parameter $q$
depends on $t$.

From the list above the guiding principle remains somewhat hidden.
There is an alternative construction by Johansson
\cite{Jo03,Jo05a} which avoids fermions altogether and  is based directly on a
determinantal weight for the line ensemble. In the {\it Addendum}
we outline a purely combinatorial scheme, which was devised by
Gessel and Viennot as based on ideas of Lindstr\"{o}m. 
All the examples from our list can be obtained through
suitable limits of the Lindstr\"{o}m-Gessel-Viennot scheme,
which therefore can be regarded as the most general 
set-up for determinantal line ensembles. 

Having the machinery of determinantal point processes at our
disposal, we turn to models of statistical physics. They are
crystals in thermal equilibrium (Section \ref{sec.3}), growth
processes (Sections \ref{sec.4}, \ref{sec.5}) and the eigenvalue
statistics of random matrices (Section \ref{sec.6}). Physical
predictions are extracted from edge scaling. We also indicate
briefly how through appropriate boundary conditions for the line
ensemble further cases of physical interest can be handled.


\subsection*{Addendum: Nonintersecting paths on directed graphs
without  loops}

We explain the Lindstr\"{o}m-Gessel-Viennot theorem in the form
stated by Stembridge \cite{St90}. One starts with a graph $(V,E)$
consisting of vertices $V$ and directed edges $E$. The graph has
no loops. A path $P$ is a sequence of consecutive vertices joined
by directed edges. $\mathcal{P}(u,v)$ denotes the set of all paths
starting at $u\in V$ and ending at $v\in V$. The paths $P$ and
$P'$ intersect, if they have a common vertex. Every edge carries a
weight $w(e)$ and every vertex a weight $\widetilde{w}(v)$. The
weight of a path $P$ is hence given by
\begin{equation}\label{2.32}
  w(P)=\prod_{e\in P\cap E}w(e)\prod_{v\in P\cap V}\widetilde{w}(v)
\end{equation}
and we set
\begin{equation}\label{2.33}
  h(u,v)=\sum_{P\in \mathcal{P}(u,v)}w(P)\,.
\end{equation}
Let us now consider an $r$-tuple $\vec{u}=\{u_1,\ldots,u_r\}$ of
starting points and an $r$-tuple $\vec{v}=\{v_1,\ldots,v_r\}$ of
end points. Let $\mathcal{P}_0(\vec{u},\vec{v})$ be the set of all
\emph{non-intersecting} $r$-tuple of paths from $\vec{u}$ to $\vec{v}$.
$\vec{u}$ and $\vec{v}$ have to be compatible, which means that any
$r$-tuple of paths in $\mathcal{P}_0(\vec{u},\vec{v})$ necessarily
connects $u_j$ to $v_j$ for $j=1,\ldots,r$. Then the
weight of $\mathcal{P}_0(\vec{u},\vec{v})$ is given by
\begin{equation}\label{2.34}
w(\mathcal{P}_0(\vec{u},\vec{v}))=\det \big(h(u_i,v_j)\big)_{1\leq
i,j\leq r}\,.
\end{equation}

\begin{figure}[t!]
  \begin{center}\hfill
  \subfigure[]{
  \psfrag{w0}[lB]{$w_0$}
  \psfrag{wp}[c]{$w_+$}
  \psfrag{wm}[c]{$w_-$}
  \includegraphics[height=5cm]{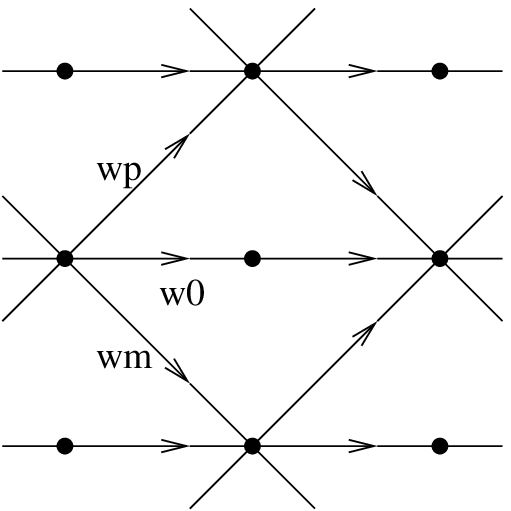}}\hfill
  \subfigure[]{
  \psfrag{1}{$1$}
  \psfrag{q+}[c]{$q_+$}
  \psfrag{q-}[l]{$q_-$}
  \includegraphics[height=5cm]{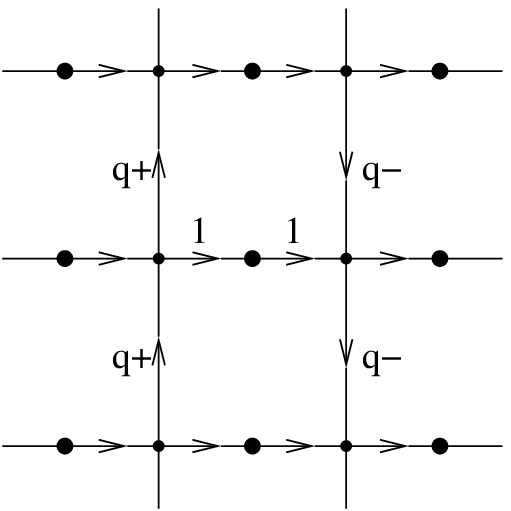}}\hfill \phantom{}
\caption{The directed graph for (a) the Aztec diamond and (b) the 3D Ising corner.}\label{Fig3}
\end{center}
\end{figure}

Let us illustrate the Gessel-Viennot scheme by a few
examples.\smallskip\\
(i) Simple random walks. Here $V=\mathbb{Z}^2$ restricted to the
even sublattice and $E$ are all nearest neighbor edges, which are
then directed either North-East or South-East. Their weight is
$\widetilde{w}(v)=1$, $w(e)=w_+$ for $e$ directed NE, and
$w(e)=w_-$ for $e$ directed SE.\smallskip\\
(ii) Aztec diamond, domino tiling \cite{Jo02,Jo05}, see Figure \ref{Fig3}(a). Here
$V=\mathbb{Z}^2$ and $E$ consists of all directed edges as in
example (i) plus edges of the form $v$ directed to $v+(1,0)$. The
horizontal edges have weight
$w_0$, the NE edges weight $w_+$, and the SE edges weight $w_-$.\smallskip\\
(iii) 3D Ising corner, lozenges tiling \cite{FS03}, see Figure \ref{Fig3}(b). Here
$V=(\mathbb{Z}\times\mathbb{Z})\cup((\mathbb{Z}+\frac{1}{2})
\times\mathbb{Z})$. The horizontal edges are between nearest
neighbors, directed East, and have weight 1. The vertical edges
are nearest neighbor for $(\mathbb{Z}+\frac{1}{2})\times\mathbb{Z}$ only. If
$\tau\in\mathbb{Z}+\frac{1}{2}$ is their 1-coordinate, then for
$\tau<0$ they are directed North and for $\tau>0$ they are
directed South with weight $q^{|\tau|}$, $0\leq q<1$.\smallskip\\
(iv) Discrete time TASEP \cite{Jo}. The setup is as in example (iii). Only
the North and South directed edges are alternating with a
$\tau$-independent weight $q \in [0,1)$. If the vertical lattice
spacing is $\varepsilon$ and the weight is $q=1-\delta\varepsilon$,
then in the limit $\varepsilon\to 0$ one obtains the one-sided
exponential jumps from item (3) above.\smallskip\\
(v) Six-vertex model at the free fermion point \cite{F05}. In the
six-vertex model, see e.g.~\cite{LW}, one draws only the South and
East pointing arrows. Because of the ice rule one then obtains a
line ensemble. However lines may touch. To achieve
non-intersecting lines, SE-edges are added but \emph{only}  for
the even sublattice. Touching is avoided by choosing the SE short
cut. There are then three weights, one for E-, SE-, and S-directed
edges, respectively. Reconstructing the six-vertex weights one
notices that they satisfy the free fermion condition. Rotating the 
space-time lattice by $\pi/4$ one arrives at Figure \ref{Fig3}(a),
only every second horizontal link is missing. They can be 
reintroduced, however, at the expense of splitting up the weight.
Therefore the Aztec diamond is equivalent to the six-vertex model at its free
fermion point.


\section{Equilibrium crystal shape}\label{sec.3}
\setcounter{equation}{0}

As a rule crystals in thermal equilibrium are faceted at low
temperatures. In this section we will discuss a simplified model,
which can be analyzed through the method of determinantal line
ensembles, see \cite{FS03,FPS04}.

\begin{figure}[t!]
\begin{center}
\psfrag{1}[][][1]{$1$}
\psfrag{2}[][][1]{$2$}
\psfrag{3}[][][1]{$3$}
\includegraphics[height=5cm]{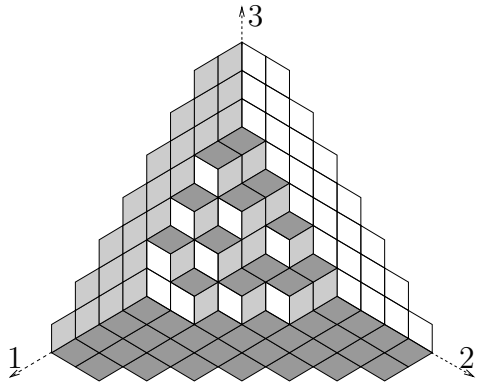}
\caption{An atom configuration of the 3D Ising corner.}\label{Fig4}
\end{center}
\end{figure}

We consider the simple cubic lattice $\mathbb{Z}^3$. Each
site can be occupied by at most one atom and the occupation variables
are denoted by $n_x=0,1$. If the nearest neighbor binding energy is $-J$, $J>0$,
then the total binding energy of all atoms is given by
\begin{equation}\label{3.1}
H=-\frac{J}{2}\sum_{|x-y|=1}n_x n_y\,.
\end{equation}
At zero temperature only configurations of minimal energy are
allowed. If exactly $N^3$ atoms are available, then they form a
cube of side-length $N$. For concreteness we assume that the cube
occupies the sites $[0,\ldots,N-1]^3\subset \mathbb{Z}^3$. If only
$N^3-M$ atoms are available, with $M<N\ll N^3$, then the binding
energy is reduced by $-3MJ$ compared to the perfect cube. However,
now there are many configurations of minimal energy. They can be
obtained by successively removing atoms from either one of the
eight corners under the constraint to cut exactly three bonds in
each step. Let us focus our attention at the corner touching the
origin, compare with Figure \ref{Fig4}.
The atoms missing at that corner can then be enumerated ed by a height
function $h(i,j)$, $i,j\geq 0$, taking only integer values such
that
\begin{equation}\label{3.2}
h(i,j)\geq 0\,,\; h(i,j)\geq h(i+1,j)\,,\; h(i,j)\geq h(i,j+1)\,,\;
\lim_{i,j \to \infty} h(i,j) = 0\,.
\end{equation}
The atoms occupy the sites $\{x\,|\,x_1=i,x_2=j,x_3\geq h(i,j)\}$. 
E.g. the perfect cube corresponds to $h(i,j) = 0$ for all $i,j$.
The number (= volume) of removed atoms is
\begin{equation}\label{3.2a}
V(h)=\sum_{i,j\geq 0}h(i,j)\,.
\end{equation}

In principle one should introduce such a height function for each
corner and the volume constraint refers jointly to all corners.
For simplicity we ignore such inessential complications, thus
disregard the other corners, and impose the volume constraint on
$h$ in the form
\begin{equation}\label{3.3}
V(h)=M\,.
\end{equation}
Every height configuration satisfying (\ref{3.2}) and (\ref{3.3})
has the same energy. Hence at zero temperature every configuration
has the same weight and the model is purely entropic. The only
task is to count.

To deal with the volume constraint it is convenient to switch to
the grand canonical ensemble as
\begin{equation}\label{3.4}
 Z_T^{-1} e^{-V(h)/T}\,.
\end{equation}
Here $T$ is the control parameter for the volume, not to be
confused with the temperature. We are interested in a macroscopic
volume, which corresponds to large $T$. Then the average volume,
average with respect to (\ref{3.4}), equals $\langle
V\rangle_T\cong T^3$ and the height is typically of order $T$. Let
us replace $h$ by $h_T$ in order to remember that the height
statistics depends on $T$. We switch from $\mathbb{Z}^3$ to
$(\mathbb{Z}/T)^3$, i.e., to a lattice spacing $1/T$ instead of 1.
In the limit $T\to \infty$ fluctuations are suppressed and one
observes a non-random macroscopic crystal shape, in formula
\begin{equation}\label{3.5}
\lim_{T\to \infty} \frac{1}{T}h_T([uT],[vT])=h_{\mathrm{ma}}(u,v)
\end{equation}
with probability one. Here $[\cdot]$ denotes the integer part,
$u,v\geq 0$, and $h_\mathrm{ma}$ is the macroscopic crystal shape.
\begin{figure}[t!]
\begin{center}
\hfill \includegraphics[width=5cm]{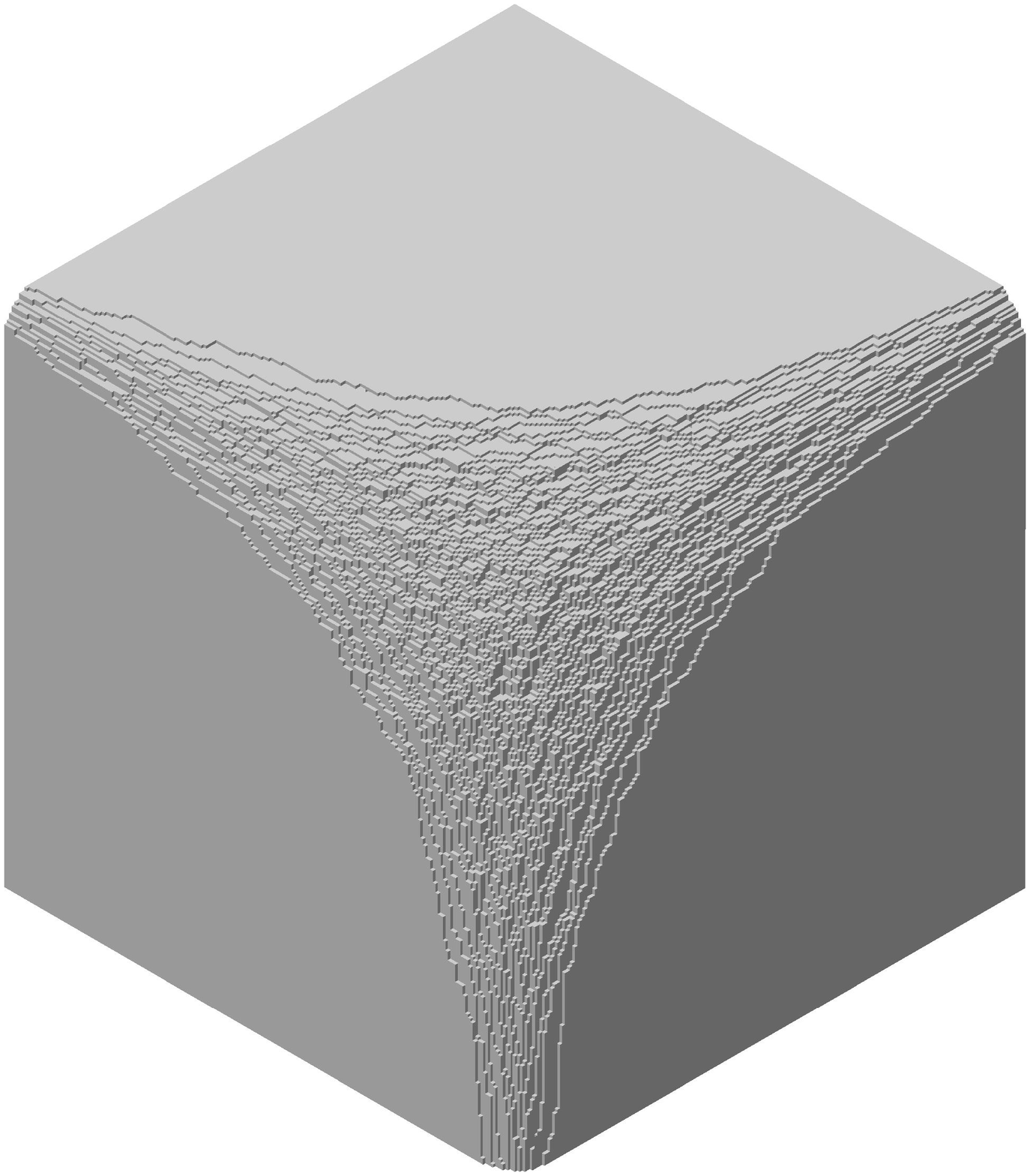}\hfill
\includegraphics[bb=0 -25 400 300,clip,width=6cm]{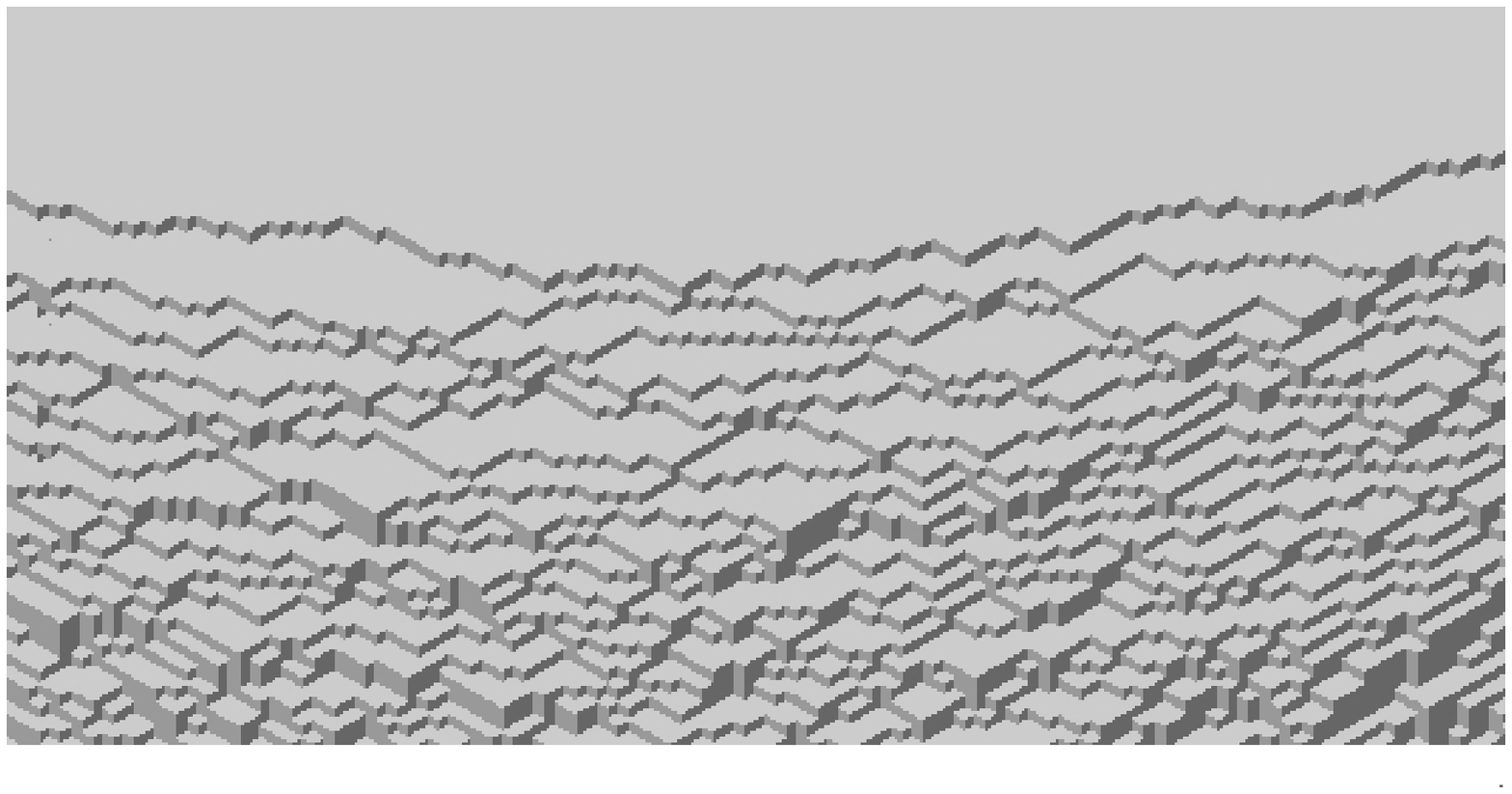}\hfill \phantom{}
\caption{Monte-Carlo simulation of the 3D Ising corner with
$M=3\times10^5$ and an enlargement close to the facet edge.}\label{Fig5}
\end{center}
\end{figure}
In Figure \ref{Fig5} we display a typical atom configuration with
volume constraint $M=3\times 10^5$. For the true crystal shape of
the low temperature Ising model with volume constraint one has to
imagine a perfect cube rounded at each corner as in Figure
\ref{Fig5}.

To have a nice looking formula for $h_\mathrm{ma}$, we choose the
volume constraint as $2\zeta_R(3)T^3$ with $\zeta_R$ the Riemann
zeta function. Let us define
\begin{equation}\label{3.6}
f(a,b,c)=\frac{1}{4\pi^2}\int^{2\pi}_0 du \int^{2\pi}_0
dv\log(a+be^{iu}+ce^{iv})\,.
\end{equation}
Then the set $S_0=\{(u,v,w)\,,\;u,v\geq0\,,\;0\leq w\leq
h_\mathrm{ma}(u,v)\}$ is parametrically given through
\begin{equation}\label{3.7}
S_0=\{2(f(a,b,c)-\log a\,,\;f(a,b,c)-\log b\,,\;f(a,b,c)-\log
c)\,|\;a,b,c>0\}\,.
\end{equation}
As expected the equilibrium shape is symmetric relative to the
(1,1,1) axis. Let $\mathcal{D}=\{(u,v)\,,\;e^{-u/2}+e^{-v/2}< 1\}$.
Then
\begin{equation}\label{3.8}
h_\mathrm{ma}(u,v)=0\quad \mathrm{on}\;\mathbb{R}^{\,2}_+ \setminus
\mathcal{D}\,.
\end{equation}
The equilibrium shape has three facets lying in the respective
coordinate planes, see Figure \ref{Fig5}. $\mathcal{D}$ is the
domain where $h_\mathrm{ma}$ is rounded. Near the facet edge, in
the direction $\tau=v-u$, one has
\begin{equation}\label{3.9}
h_\mathrm{ma}(r,\tau)=\frac{2}{3}\cosh(\tau/4)\pi^{-1}2^{1/4}r^{3/2}\,,
\end{equation}
valid for small $r$, where $r$ denotes the distance away from the
edge. The 3/2-exponent is known as Pokrovsky-Talapov law
\cite{PT}.

The expression (\ref{3.6}) has a simple physical meaning, for
which we switch to the coordinate frame with (1,1,1) as 3-axis. As
can be seen from Figure \ref{Fig4} the along (1,1,1) projected
height profile yields a perfect tiling of the plane with lozenges
which are oriented either with angle 0, or $2\pi/3$, or $4\pi/3$.
Conversely, a tiling by lozenges such that asymptotically in each
of the three segments there is only a single type translates back
to an admissible height configuration. We now focus our attention
on a small neighborhood of a point in $\mathcal{D}$. For large $T$
the curvature can be ignored and under projection the tiling is
such that the fraction of each type of lozenges remains fixed.
Again the grand canonical version is easier to control and we
assign to the three types of lozenges the Boltzmann weights
$a,b,c$, respectively. $f(a,b,c)$ from (\ref{3.6}) is the free
energy of such a tiling. Note that $f(\lambda a,\lambda b, \lambda
c)=f(a,b,c)+\log\lambda$, as it has to be. A tiling of the plane
corresponds to a flat surface with a non-random slope determined
through $a,b,c$ and $f(a,b,c)$ takes the role of the surface
tension.

Switching back to the original coordinate frame by this
construction one obtains the surface tension $\sigma$ depending on
the macroscopic slope $\nabla h$ of the height function. The
explicit formula is unwieldy and not so instructive. We now give
ourselves some macroscopic height profile $h$ defined on
$(\mathbb{R}_+)^2$. It must satisfy $h\geq 0$ and $\partial_1
h\leq 0$, $\partial_2 h\leq 0$. In the limit $T\to \infty$ the
macroscopic free energy is additive and hence the prescribed
profile $h$ has the total free energy
\begin{equation}\label{3.10}
\mathcal{F}(h)=\int_{\mathbb{R}_+^{\,2}} dx_1 dx_2 \sigma(\nabla
h(x_1,x_2))\,.
\end{equation}
Minimizing $\mathcal{F}(h)$ over admissible height profiles and
under the constraint of constant volume,
\begin{equation}\label{3.11}
\int_{\mathbb{R}_+^{\,2}}dx_1 dx_2 h(x_1,x_2)=2\zeta_R(3)\,,
\end{equation}
yields the actual height profile $h_\mathrm{ma}$ as implicitly
defined in (\ref{3.7}). Here $\zeta_R$ is the Riemann zeta
function.

Our real interest are the small shape fluctuations on top of the
macroscopic profile. The three facets are perfectly flat, no
fluctuations. For the rounded piece one can use the standard
Einstein fluctuation argument, which means to expand
$\mathcal{F}(h_\mathrm{ma}+\delta h)-\mathcal{F}(h_\mathrm{ma})$
to second order in $\delta h$,
\begin{equation}\label{3.12}
\mathcal{F}(h_\mathrm{ma}+\delta
h)-\mathcal{F}(h_\mathrm{ma})\cong \int_{\mathbb{R}_+^{\,2}}dx
\int_{\mathbb{R}_+^{\,2}}dx'\nabla \delta h(x)\cdot\,\mathrm{Hess}\,\sigma(\nabla
h_{\mathrm{ma}}(x_1,x_2))\nabla\delta h(x')\,,
\end{equation}
where Hess\,$\sigma$ is the $2\times 2$ matrix of second derivatives
of $\sigma$ with respect to $\nabla h$.  The inverse of the
operator appearing in the quadratic form for $\delta h(x)$ defines the
covariance matrix $C(x,x')$. The assertion is that, for large $T$,
\begin{equation}\label{3.13}
h_T([uT],[vT])-Th_\mathrm{ma}(u,v)\,,\quad u,v\in \mathcal{D}\,,
\end{equation}
become jointly Gaussian with covariance matrix $C$. In fact, as
proved in \cite{Ke}, such a property holds provided one integrates
(\ref{3.13}) against a smooth test function depending on $u,v$.
Roughly, $C$ is the covariance of a free massless Gaussian field
with a strength which is modulated by $h_\mathrm{ma}$. Note that
the fluctuations are only $\mathcal{O}(1)$, thus tiny compared to
the same number of independent random variables which would amount
to a size $(\sqrt{T})^2$. If in (\ref{3.13}) one integrates over a
small square in $u,v$, then this spacially averaged height has
fluctuations of size $\log T$.

We have left out the most intriguing fluctuations close to the
facet edge. There the crystal steps have much more freedom to
fluctuate as compared to the steps in the disordered zone, which are squeezed
by their neighbors. To be able to analyse facet edge fluctuations,
we have to set up the line ensemble.

\begin{figure}[t!]
 \begin{center}
  \subfigure[]{\psfrag{1}[][][1]{$1$}
  \psfrag{2}[][][1]{$2$}
  \psfrag{3}[][][1]{$3$}
  \includegraphics[width=0.45\textwidth]{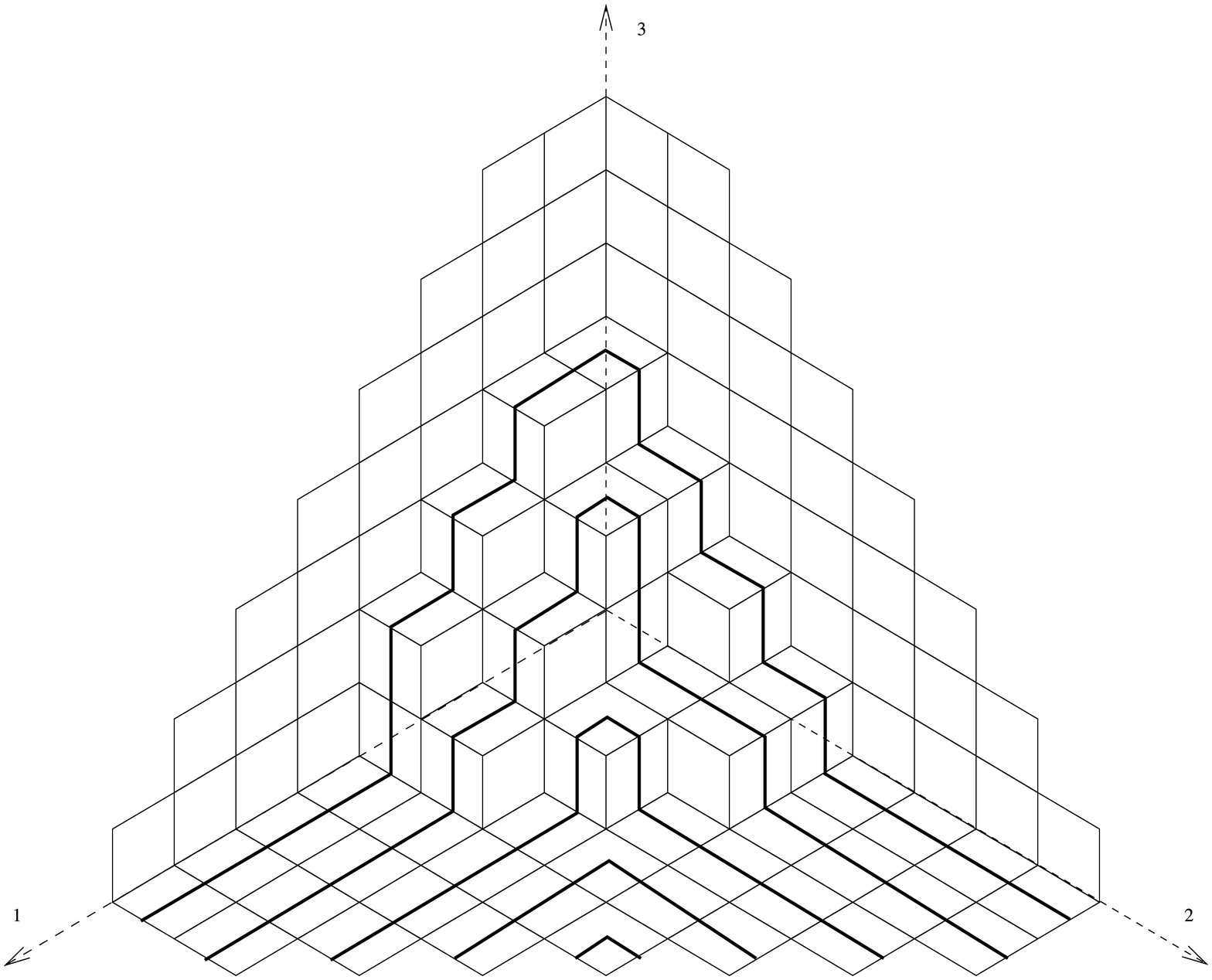}}\hfill
  \subfigure[]{\psfrag{t}[][][1]{$t$}
  \psfrag{j}[][][1]{$j$}
  \psfrag{1}[][][1]{$1$}
  \psfrag{h0}[][][1]{$h_0$}
  \psfrag{h1}[][][1]{$h_1$}
  \psfrag{h2}[][][1]{$h_2$}
  \psfrag{h3}[][][1]{$h_3$}
  \includegraphics[width=0.45\textwidth]{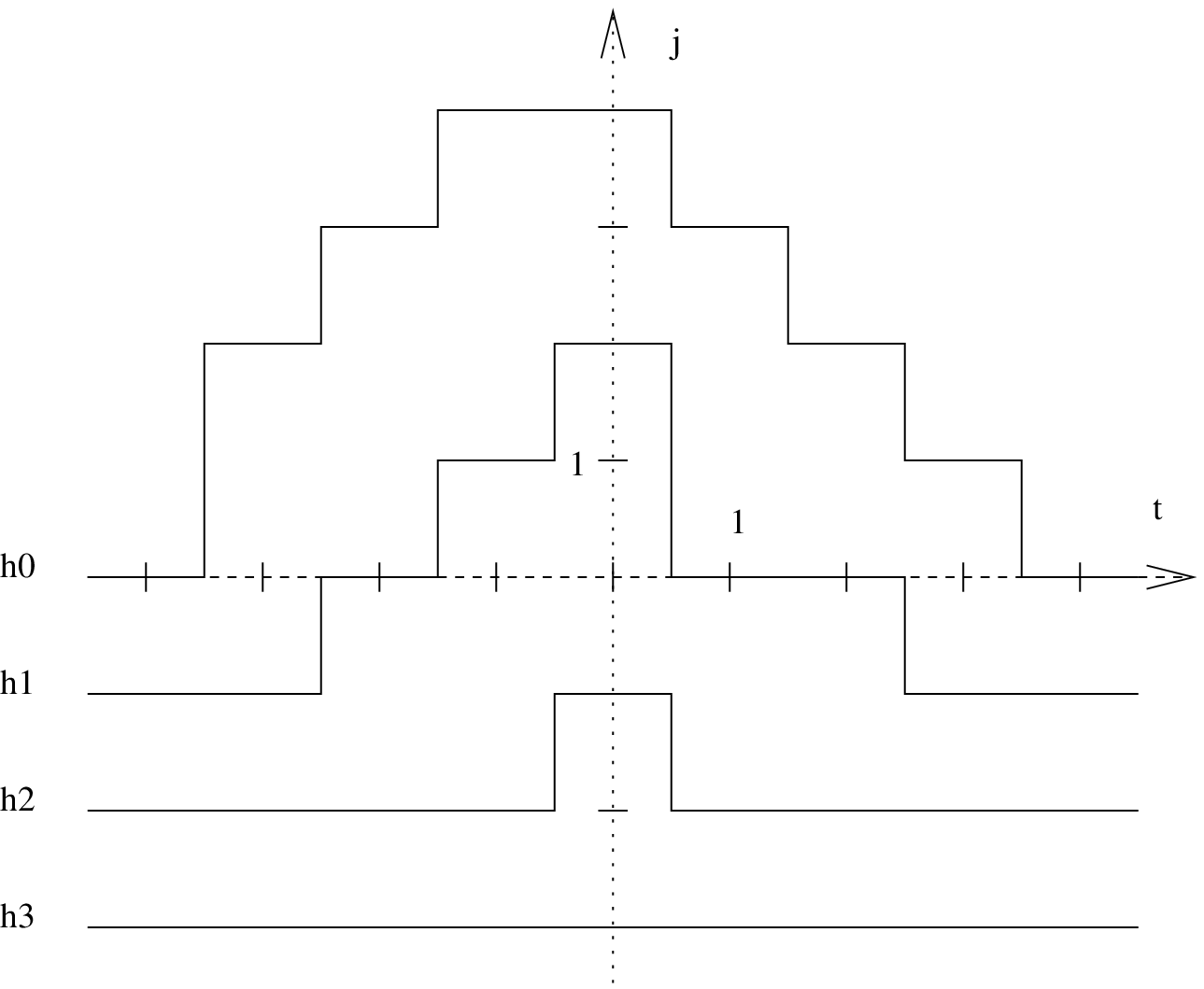}}
\caption{The gradient lines of the height profile from Figure
\ref{Fig4}:  (a) the $(111)$-projection, (b) as line ensemble.}\label{Fig6}
\end{center}
\end{figure}

We return to (\ref{3.2}) and use instead of $h$ the gradient lines
$h_\ell(t)$, $t\in\mathbb{Z}$, $\ell=0,-1,\ldots$, see Figure \ref{Fig6}(a). They are
defined through
\begin{equation}\label{3.14}
  t=j-i\,,\quad h_\ell(t)=h(i,j)+\ell(i,j)\,,
\end{equation}
where
\begin{equation}\label{3.15}
\ell(i,j)=-(i+j-|i-j|)/2\,.
\end{equation}
Then
\begin{equation}\label{3.16}
h_\ell(t)\leq h_\ell(t+1)\,,\;t<0\,,\;h_\ell(t)\geq
h_\ell(t+1)\,,\; t\geq 0\,,
\end{equation}
with the asymptotic condition
\begin{equation}\label{3.17}
\lim_{|t|\to\infty}h_\ell(t)=\ell\,.
\end{equation}

We extend $h_\ell$ to a piecewise constant function on
$\mathbb{R}$ such that the jumps are at the midpoints, i.e., at
some point of $\mathbb{Z}+\frac{1}{2}$.
The gradient lines are then non-intersecting, in the sense that
\begin{equation}\label{3.18}
h_{\ell-1}(t)<h_\ell(t)\,,\quad t\in\mathbb{R}\,,
\end{equation}
compare with Figure \ref{Fig6}(b).
For the $\ell$-th line let $t_{\ell,1}<\ldots<t_{\ell,k(\ell)}<0$
be the times of up-steps with step sizes
$s_{\ell,1},\ldots,s_{\ell,k(\ell)}$ and let
$0<t_{\ell,k(\ell)+1}<\ldots<t_{\ell,k(\ell)+n(\ell)}$ be the
times of down-steps with step sizes
$-s_{\ell,k(\ell)+1},\ldots,-s_{\ell,k(\ell)+n(\ell)}$. The volume
under the height function $h$ is the sum over the ``areas of excitation''
for each line. Dividing these areas horizontally results in
\begin{equation}\label{3.19}
V(h)=\sum^0_{\ell=-\infty}\sum^{k(\ell)+n(\ell)}_{j=1}s_{\ell,j}
|t_{\ell,j}|\,.
\end{equation}
Therefore the Boltzmann weight of a line configuration is
\begin{equation}\label{3.20}
\prod^0_{\ell=-\infty}\exp\big[-\frac{1}{T}\sum^{k(\ell)+n(\ell)}_{j=1}
s_{\ell,j}|t_{\ell,j}|\big]\,.
\end{equation}

To prove that (\ref{3.20}) defines a determinantal process we use
the directed graph from example (iii) of the addendum to Section
\ref{sec.2}. The vertices of the graph are
$(\mathbb{Z}\cup(\mathbb{Z}+\frac{1}{2}))\times\mathbb{Z}$. The
horizontal bonds are between nearest neighbors and directed to the
right. The vertical bonds are only on
$(\mathbb{Z}+\frac{1}{2})\times\mathbb{Z}$. For positive $t$ they
are directed downwards, for negative $t$ they are directed
upwards. To every horizontal bond we assign the weight one. To
every vertical bond with 1-coordinate $t+\frac{1}{2}$, $t$
integer, we assign the weight
\begin{equation}\label{3.21}
q_t=q^{|t+\frac{1}{2}|}\,,\quad q=e^{-1/T}\,.
\end{equation}
A path on this directed graph has a weight which is given
by the product of weights for each step. The line ensemble
$\{h_\ell,\ell=0,-1,\ldots\}$ is a collection of non-intersecting paths on
this graph and their weight agrees with (\ref{3.20}).

Following the scheme of Section \ref{sec.2} we introduce the variables
$\eta(j,t)$ in such a way that
\begin{equation}\label{3.22}
\eta(j,t)=
  \begin{cases}
    1, & \text{if there is a line passing through}\;(t,j)\in \mathbb{Z}^2 \,,\\
   0 & \text{otherwise}.
  \end{cases}
\end{equation}
From the construction of the addendum to Section \ref{sec.2} we
know that $\eta(j,t)$ has determinantal moments. The covariance
kernel follows from (\ref{2.33}). Let us first consider a single
up-step with weight $q<1$. According to (\ref{2.33}) a single path
starting at $j$ and being at $i$ one time unit later has the weight
\begin{equation}\label{3.23}
\big(\mathfrak{t}_+(q)\big)_{ij}=
  \begin{cases}
    q^{i-j} & \text{for }i-j\geq 0\,, \\
    0 & \text{for }i-j<0\,.
  \end{cases}
\end{equation}
The up-step transfer matrix $T_+(q)$ for a particle configuration
at an integer column to the next one  is then the second
quantization of $\mathfrak{t}_+(q)$, i.e., $T_+(q)$ restricted to
the $n$-particle space equals
$S_\mathrm{a}\mathfrak{t}_+(q)\otimes\ldots\otimes
\mathfrak{t}_+(q)S_\mathrm{a}$, i.e., the anti-symmetrized
$n$-fold product.

The same transfer matrix can be obtained also from direct summation.
Initially there are $n$ points. They move upwards
under the  non-crossing constraint. We want to compute the Boltzmann
weight $\langle y_1,\ldots,y_n\,|T_+(q)|\,x_1,\ldots,x_n\rangle$
for initial  configuration
$(x_1,\ldots,x_n)=(x)_n$ and final configuration
$(y_1,\ldots,y_n)=(y)_n$. If $(x)_n=(y)_n$, i.e., no step at all,
one has the contribution 1 for $T_+(q)$. If $(y)_n$ differs from
$(x)_n$ by a single step, one has the contribution
\begin{equation}\label{3.23a}
-q\sum_{k\in\mathbb{Z}}a_k a^\ast_{k+1}
\end{equation}
for $T_+(q)$, where the minus sign arises from the chosen order of Fermi operators.
 Similarly a difference of two steps results in the
contribution
\begin{equation}\label{3.24}
q^2\frac{1}{2}\sum_{k_1,k_2\in\mathbb{Z}}a_{k_1} a_{k_2}
a^\ast_{k_2+1} a^\ast_{k_1+1}\,.
\end{equation}
Therefore
\begin{equation}\label{3.25}
T_+(q)=\sum^\infty_{n=0} \frac{(-q)^n}{n!}
\sum_{k_1,\ldots,k_n\in\mathbb{Z}}a_{k_1} \ldots
a_{k_n}a^\ast_{k_n+1}\ldots a^\ast_{k_1+1}\,.
\end{equation}
Using properties of Schur polynomials the sum can be carried out
resulting in
\begin{equation}\label{3.26}
T_+(q)=\exp\big[\sum_{i,j\in\mathbb{Z}}a^\ast_i
\mathfrak{g}_+(q)_{ij}a_j\big]\,,
\end{equation}
where
\begin{equation}\label{3.27}
(\mathfrak{g}_+(q))_{ij}=q^{i-j}\frac{1}{i-j}\theta(i-j-1)
\end{equation}
with $\theta(j)=1$ for $j\geq 0$ and  $\theta(j)=0$ for $j<0$,
which is in agreement with the previous argument. Note that
$\mathfrak{g}_+(q)$ is not symmetric because of one-sided steps.

By the same argument, for down-steps only, $T_-(q)$ is the second
quantization of $\mathfrak{t}_-(q)=\mathfrak{t}_+(q)^\ast$ which
implies $T_-(q)=T_+(q)^\ast$.

 With this result the Boltzmann weight from $t$ to $t+1$ is
$T_+(q_t)$ for $t\leq -1$ and $T_-(q_t)=T_+(q_t)^\ast$ for $t\geq
0$. In the classification of Section \ref{sec.2} the generator of
the time propagation is time-dependent.

To obtain the covariance kernel for the point process $\eta(j,t)$
two limit procedures are still needed. Firstly we let exactly
$M+1$ lines run from $t=-S$ to $t=S$ and require that at $\pm S$
the sites $[0,-1,\ldots,-M]$ are occupied. The line ensemble of
interest is recovered in the limits $M\to\infty$ and $S\to\infty$.
The formula for the covariance kernel can be found in \cite{Fth},
Eq.~(5.39).

Independent of this specific formula, the line ensemble has a
rather striking appearance. For large $T$, the plane is divided
into an ordered and disordered zone which is bordered by the two
lines
\begin{equation}\label{3.28}
  b^+_\infty(t/T)=-2T\log(1-e^{-|t|/2T})\,,\quad
  b^-_\infty(t/T)=2T\log(1-\tfrac{1}{2}e^{-|t|/2T})\,.
\end{equation}
In the ordered zone, with large probability, $\eta(j,t)=0$ for
$j\geq b^+_\infty(t/T)+\mathcal{O}(T^{1/3})$ and $\eta(j,t)=1$ for
$j\leq b^-_\infty(t/T)-\mathcal{O}(T^{1/3})$. The width of the
transition region between ordered and disordered is
$\mathcal{O}(T^{1/3})$, but to find out requires a detailed asymptotic
analysis.

Let us now focus our attention on a point $([uT],[vT])$ of fixed
relative location inside the disordered zone,
i.e.,~$b^-_\infty(u)<v<b^+_\infty(u)$. For $T\to\infty$, close to
this point, the line statistics becomes stationary in space-time.
It encodes the statistics of the tiling of the plane with lozenges
at a fixed relative fraction depending on the reference point
$(u,v)$ through $\nabla h_{\mathrm{ma}}(u,v)$. On a mesoscopic
scale $\eta(j,t)$ is averaged over regions of linear size $T$, but
still inside the disordered zone. One then recovers the Gaussian
shape fluctuations for the rounded piece of $h_{\mathrm{ma}}$, as
discussed before, see (\ref{3.12}).

Clearly, the facet edge corresponds to the top line $h_0(t)$.
$h_0(t)$ has more space to fluctuate. Thus its fluctuation
behavior is expected to be very different from the lines deep
inside the disordered zone. $h_0(t)$ is the microscopic edge
between ordered and disordered. The properly adjusted scaling with
$T$ is thus referred to as ``edge scaling'', which will be
explained
 in Section \ref{sec.8}.


\section{Growth models in one dimension: PNG}\label{sec.4}
\setcounter{equation}{0}

For the Ising corner the appropriate non-intersecting line ensemble
can be seen by inspection. Still, it is a sort of miracle that the
physical Boltzmann weight makes the line ensemble determinantal.
For growth processes the line ensemble is much more hidden and to
bring it to light is one part of the discoveries over the recent
years. Of course, the construction works only for very special
growth processes. Also, the method is restricted to one dimension.
Even then it does not yield information on temporal correlations.

To stress the similarity with the Ising corner we consider in this
section the polynuclear growth (PNG) model in the droplet
geometry. We use $x\in \mathbb{R}$ for physical space and $T$,
$T\geq 0$, for the growth time, which should not be confused with
the time for the line ensemble. The PNG model describes the
stochastic evolution of the height profile $h(x,T)$, which takes
integer values only. A point $x$ where
$h(x+\varepsilon,T)-h(x-\varepsilon,T)=1$ , $\varepsilon$ small,
is referred to as up-step, while
$h(x+\varepsilon,T)-h(x-\varepsilon,T)=-1$ is a down-step. Larger
steps do not occur. The height profile evolves by two mechanisms.
Firstly, up-steps move to the left with velocity $-1$ and
down-steps to the right with velocity $+1$. Physically the idea is
that material can easily attach once a step is formed. Steps may
collide, upon which they simply coalesce. Coalescence should be
thought of as a damping, or smoothening, mechanism which, as in
any other nonequilibrium system, has to be counterbalanced  by a
suitable driving force. For PNG it is given through random
nucleation events. They have Poisson statistics in space-time. At
a nucleation event a nearby pair of an up-step and a down-step is
created, which then move apart according to the deterministic
rule.

In the droplet geometry, one imposes initially $h(x,0)=0$ with a
single nucleation event at $(x,t)=(0,0)$. The droplet constraint
means that nucleation is allowed only on the layers with $\{h\geq
1\}$. Clearly, the height will grow faster in the center than at
the edges $x=\pm T$. In fact, for nucleation intensity 2, one has
\begin{equation}\label{4.1}
\lim_{T\to\infty}\frac{1}{T}h(yT,T)=2\sqrt{1-y^2}\,,\quad |y|\leq
1\,,
\end{equation}
with probability one. Thus the macroscopic growth shape is a
droplet, which explains the name.

\begin{figure}[t!]
\begin{center}
\psfrag{x}{$x$}
\psfrag{t}{$t$}
\psfrag{t=T}{$t=T$}
\psfrag{h0}[c]{$h=0$}
\psfrag{h1}[c]{$h=1$}
\psfrag{h2}[c]{$h=2$}
\psfrag{h3}[c]{$h=3$}
\includegraphics[height=5cm]{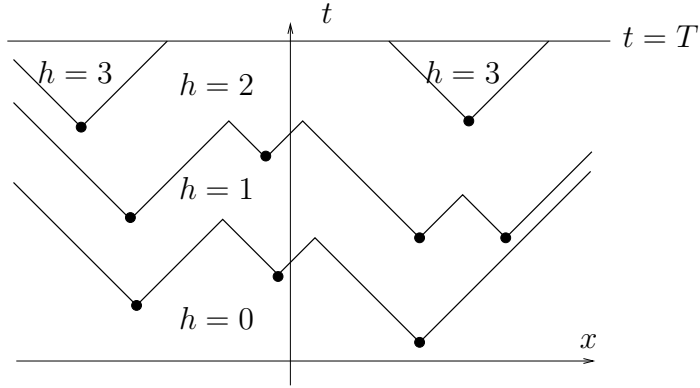}
\caption{Nucleation events, world lines of steps, and the
associated height profile.}\label{Fig7}
\end{center}
\end{figure}

Since $h(x,T) \in  \mathbb{Z}$, the height profile is determined
by the positions of the up- and down-steps and it is sometimes
convenient to switch to the step world lines. We use the
relativistic convention according to which the $t$-axis points
upwards. Steps have speed 1 (the speed of light). Nucleation
events lie in the forward light cone $\{(x,t)\;|\;|x|\leq t\}$ of
the origin only and are Poisson distributed with intensity 2. Each
nucleation event is the apex of a forward light cone,
corresponding to the world lines of the there created pair of an
up- and a down-step. The world lines annihilate each other upon
collision, where the annihilation events have to be determined
sequentially starting at $t=0$. As a result one obtains an
ensemble of broken lines. They divide the forward light cone
$\{(x,t)\;|\;|x|\leq t\}$ of the origin into layers of constant
height. The lowest layer has height 1, since there is a Poisson
point at $(0,0)$. Crossing the broken line the height increases to
2, {\it etc}.. In Figure \ref{Fig7} we display an example with
flat initial conditions $h(x,0)=0$. The height profile $h(x,T)$ at
time $T$ records the height along the horizontal line $t=T$.

The construction of the line ensemble is based on the observation
that at each coalescence of two steps one loses information, since
there are many ways how a particular height profile could have
been achieved. To retain the information we set $h_0(x,T)=h(x,T)$,
the PNG profile, and introduce at time $T=0$ the extra
book-keeping heights $h_\ell(x,0)=\ell$, $\ell=-1,-2,\ldots$. By
definition $h_0(x,T)$ evolves according to the PNG rules. In
addition, whenever there is a coalescence event at line $\ell$, it
is instantaneously copied as a nucleation event at the same
location to the lower lying line $\ell-1$. Random nucleation
events occur only at the top line $h_0$. For the book-keeping
heights $h_\ell$, $\ell\leq -1$, the steps move deterministically
and coalesce according to the PNG rule.

As before the occupation variables $\eta_T(j,t)$, $|t|\leq T$,
$j\in\mathbb{Z}$, are defined through
\begin{equation}\label{4.2}
\eta_T(j,t)=
  \begin{cases}
    1\,, & \text{if there is a line with}\;h_\ell(t,T)=j\,, \\
    0 & \text{otherwise}.
  \end{cases}
\end{equation}
The pattern for $\eta_T(j,t)$ has an appearance rather similar to
the Ising corner. There is a disordered zone sharply separated
from the ordered zone. Above the droplet, $j>2\sqrt{T^2-t^2}$, in
essence, $\eta_T(j,t)=0$, while below, $j<-2\sqrt{T^2-t^2}$, one has
$\eta_T(j,t)=1$. At the two borders, $\eta_T(j,\pm T)=1$ for
$j\leq 0$ and $\eta_T(j,\pm T)=0$ for $j>0$, see Figure
\ref{Fig1}(b).

The joint distribution of the line ensemble
$\{h_\ell(t,T)\,,\;\ell\in\mathbb{Z}_-\}$, $T$ fixed, is
determined dynamically through the PNG rules.
Surprisingly enough, precisely the same distribution can be  generated also
statically. For this purpose let us consider a family
$\{x_\ell(t)\,,\;\ell\in\mathbb{Z}_-\}$ of independent, time
continuous random walks on $\mathbb{Z}$, i.e., $x_\ell(t)$ takes
values in $\mathbb{Z}$. We require that $x_\ell(\pm T)=\ell$. The
random walks jump to nearest neighbor sites only and do so with
rate 1. In other words, the right and left jumps occur
independently at Poisson times with rate 1. $x_\ell(-T)=\ell$
and the random walk is constrained to arrive at $\ell$ at time
$t=T$. These random walks are conditioned not to intersect.
The conditioned walks are denoted by $\widetilde{x}_\ell(t)$. Then
 $h_\ell(t,T)=\widetilde{x}_\ell(t)$ jointly in
distribution. The proof is not difficult, but requires some
notation. We refer to \cite{PS} for the details. From the static
construction, it is obvious that $\eta_T(j,t)$ has determinantal
moments. In fact the correlation kernel has a structure simpler
than the one for the Ising corner. Let us first consider the point
process $\{\eta_T(j,0)$, $j\in\mathbb{Z}\}$, along the line $t=0$.
We introduce the one-particle operator
\begin{equation}\label{4.3}
(\mathfrak{h}_T\psi)_j=-\psi_{j+1}-\psi_{j-1}
+\frac{j}{T}\psi_j\,.
\end{equation}
The eigenvalue equation for $\mathfrak{h}_T$,
$\mathfrak{h}_T\psi^{(\lambda)}=\lambda\psi^{(\lambda)}$, has
eigenvalues $\lambda=\frac{m}{T}$, $m\in\mathbb{Z}$, and
eigenvectors $\psi^{(\lambda)}_j=J_{j-m}(2T)$, where $J_n(z)$ is
the Bessel function of integer order $n$ \cite{Ab}. The
correlation kernel $B_T$ for $\{\eta_T(j,0)$, $j\in\mathbb{Z}\}$
is given by
\begin{equation}\label{4.4}
B_T(i,j)=\sum_{m\leq 0}J_{i-m}(2T)J_{j-m}(2T)\,,
\end{equation}
also known as discrete Bessel kernel. The distribution of
$\eta_T(j,0)$, $j\in\mathbb{Z}$, equals the positional
distribution for the ground state of an ideal Fermi gas on the
one-dimensional lattice $\mathbb{Z}$ with nearest neighbor hopping
and a linear potential of slope $1/T$. The first particle is
located typically at $j=2T$, while the last hole sits near to
$j=-2T$.

The extension to fermionic time uses the one-particle Hamiltonian
\begin{equation}\label{4.5}
(\mathfrak{g}\psi)_j=-\psi_{j+1}-\psi_{j-1} +2\psi_j
\end{equation}
which, up to the overall minus sign, is the generator for the
time-continuous random walk $x_\ell(t)$. Then the space-time
correlation kernel is
\begin{equation}\label{4.6}
B_T(j,t;j',t')=
  \begin{cases}
(e^{-t\mathfrak{g}}R_T e^{t'\mathfrak{g}})_{jj'}& \text{for }t\geq t'\,, \\
 -(e^{-t\mathfrak{g}}(1-R_T) e^{t'\mathfrak{g}})_{jj'} & \text{for
 }t<t'\,,
  \end{cases}
\end{equation}
with $|t|,|t'|\leq T$.

\begin{figure}[t!]
\begin{center}
\includegraphics[height=5cm]{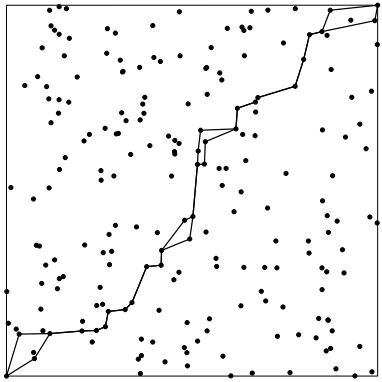}
\caption{Optimal leftmost and rightmost directed polymer over Poisson points.}\label{Fig8}
\end{center}
\end{figure}

Already in their seminal paper \cite{KPZ} Kardar, Parisi, and Zhang
 recognized that growth processes can be
reformulated as a directed polymer in a random potential, which
gives the subject an equilibrium statistical mechanics flavor.
This suggests that the PNG model, hopefully also the associated
line ensemble, must have a transcription to directed polymers. The
hint comes from the space-time picture of the step world lines.
According to convention, the space-time diagram is rotated by
$-\pi/4$. Then the Poisson points
$\omega=\{\omega_j,j=1,2,\ldots\}$ lie then in the positive
quadrant $(\mathbb{R}_+)^2$ of the plane. The broken lines are
parallel to the coordinate axes and have roughly a hyperbolic
shape. The directed polymer $\gamma$ is, so to speak, dual to the
broken lines. $\gamma$ starts at the origin $(0,0)$ and ends at
$(u,v)\in(\mathbb{R}_+)^2$. $ \gamma$ consists of consecutive line
segments, which have Poisson points (and $(u,v)$) as their
end points. $\gamma$ is directed in the sense that each line
segment must have positive slope. In other words, if $\gamma$ is
at the Poisson point $\omega_j$, then the next Poisson point of
$\gamma$ must be in the forward light cone with
apex $\omega_j$, see Figure \ref{Fig8}.

To each directed polymer $\gamma$ one associates the ``energy''
\begin{equation}\label{4.7}
  E(\gamma)=\textrm{number of Poisson points along}\; \gamma\,.
\end{equation}
As can be seen from the geometrical construction, the height is
just the energy of an optimal path. Thus we define
\begin{equation}\label{4.8}
  e(u,v)=\max_{\gamma:(0,0)\to(u,v)} E(\gamma)\,,
\end{equation}
where the maximum is over all directed polymers from $(0,0)$ to
$(u,v)$. In general, there are many maximizing paths. From the
geometric construction it is obvious that
\begin{equation}\label{4.9}
h(x,T)=e(T-x,T+x)\,,
\end{equation}
up to a scale factor $\sqrt{2}$ which is compensated by demanding that
the Poisson points for the directed polymer have density 1.
The dynamical PNG model is replaced by finding the energy
of an optimal directed path in a random potential. If we would
associate to $\gamma$ the Boltzmann weight $e^{\beta E(\gamma)}$,
then the optimization problem corresponds to the zero temperature
limit $\beta\to\infty$.

Since $h_0 \equiv h$, through the directed polymer we have
reconstructed the top line of the line ensemble. To find out
$h_{-1}$, we recall the space-time picture of nucleation and
coalescence events. Let us call $\omega=\omega^{(0)}$ the original
Poisson points and $\omega^{(1)}$ the corresponding coalescence
points. Of course, $\omega^{(1)}$ will no longer be Poisson
distributed. We now regard $\omega^{(1)}$ as the second generation
nucleation events and construct from them $h_{-1}$ by the same
rules as we did $h_0$ from $\omega^{(0)}$. In turn the coalescence
points $\omega^{(2)}$ of $\omega^{(1)}$ are regarded as nucleation
points for $h_{-2}$, {\it etc}.. If $T$ is fixed, eventually no
points remain and from thereon $h_\ell(x,T)=\ell$.


\section{Growth models in one dimension: TASEP}\label{sec.5}
\setcounter{equation}{0}

A second popular growth model is the TASEP (totally asymmetric
simple exclusion process). Its height function $h(j,T)$,
$j\in\mathbb{Z}$, $T\geq 0$, takes values in $2\mathbb{Z}$ for $j$
even and in $2\mathbb{Z}+1$ for $j$ odd. The height differences
are 1 in absolute value, $|h(j+1,T)-h(j,T)|=1$. In the growth
dynamics local minima of $h$ are increased independently by two
units after an exponentially distributed waiting time. More
precisely, if $j_\mathrm{m}$ is a local minimum of $h$ at time
$T$, then $h(j_\mathrm{m},T)$ is updated to
$h(j_\mathrm{m},T+t_\mathrm{w})= h(j_\mathrm{m},T)+2$ with
$t_\mathrm{w}$ the independent waiting time. If thereby a new
local minimum is created, one assigns to it a further independent
waiting time, {\it etc}..

The name TASEP comes from interpreting the difference
$\eta_T(j)=\big(1-(h(j+1,T)-h(j,T))\big)/2$ as occupation
variables, where $\eta_T(j)=0$ refers to site $j$ empty and
$\eta_T(j)=1$ to site $j$ occupied by a particle. Translating our
updating rule, each particle jumps to the right after an
independent exponentially distributed waiting time provided the
right neighbor site is empty. ``Exclusion'' means that there is at
most one particle per site and ``totally asymmetric simple''
refers to nearest neighbor jumps exclusively to the right. One
could modify the model to its partially asymmetric version by
allowing also jumps to the left, respecting exclusion. In the
growth interpretation some material would detach from the surface.

To construct the line ensemble we consider the particular initial
condition
\begin{equation}\label{5.1}
  h(j,0)=|j|\,,
\end{equation}
which is the analogue of the droplet for PNG. In the course of
time the cone fills up and
\begin{equation}\label{5.2}
\lim_{T\to\infty}\frac{1}{T} h([uT],T)=h_{\mathrm{ma}}(u)\,,
\end{equation}
where
\begin{equation}\label{5.3}
h_{\mathrm{ma}}(u)=
\begin{cases}
    |u| & \text{for}\;|u|\geq 1\,, \\
    \frac{1}{2}(u^2+1) & \text{for}\;|u|\leq 1\,.
  \end{cases}
\end{equation}
The key for the construction comes from the directed polymer. Let
us consider the positive quadrant $(\mathbb{Z}_+)^2$ and attach to
each site $(i,j)$ the random variable $w(i,j)$. The $w(i,j)$'s are
independent and have a unit exponential distribution. They are
linked to the waiting times in the growth steps. As before, we
introduce a lattice path $\gamma$. It starts at $(1,1)$, ends at
$(m,n)$, and at each step it can either move East or North. To
such an East-North directed path $\gamma$ we associate the energy
\begin{equation}\label{5.4}
E(\gamma)=\sum_{(i,j)\ni\gamma} w(i,j)\,.
\end{equation}
The energy of an optimal path is then
\begin{equation}\label{5.5}
G(m,n)=\max_{\gamma:(1,1)\to(m,n)} E(\gamma)\,.
\end{equation}
$G(m,n)$ is related to the TASEP height through
\begin{equation}\label{5.6}
\mathbb{P}(\{G(m,n)\leq T\})= \mathbb{P}(\{m+n\leq h(m-n,T)\})\,.
\end{equation}

In the spirit of the PNG model one reinterprets $G(m,n)$ as the
height of yet another growth process $\widetilde{h}(j,\tau)$ by
setting
\begin{equation}\label{5.6a}
\widetilde{h}(j,\tau)=G(\tau-1+j,\tau-1-j)\,,\quad |j|<\tau-1\,.
\end{equation}
Hence $j\in\mathbb{Z}$, $\tau$ is the discrete growth time, and
$\widetilde{h}(j,\tau)\in \mathbb{R}$. The growth process is
defined through the stochastic iteration
\begin{eqnarray}\label{5.7}
\widetilde{h}(j,0)&=&0, \\
\widetilde{h}(j,\tau+1)&=& \left\{
\begin{array}{ll}
 \max\{\widetilde{h}(j-1,\tau),\widetilde{h}(j+1,\tau)\}& \\
 \phantom{\max}\hspace*{10pt}+w((\tau+j)/2,(\tau-j)/2), & \textrm{if }(-1)^{j+\tau}=1,\\[6pt]
 \widetilde{h}(j,\tau), & \textrm{if }(-1)^{j+\tau}=-1,
\end{array}\right. \nonumber \\
& & \phantom{0,\,}\textrm{ for }|j|<\tau+1,\nonumber \\
\widetilde{h}(j,\tau+1)&=&0 \textrm{ for
}|j|\geq\tau+1\,.\nonumber
\end{eqnarray}
From (\ref{5.2}) one infers that for large $\tau$
\begin{equation}\label{5.7a}
\widetilde{h}(j,\tau)\cong\frac{2\tau}{1+(j/\tau)^2}\,,\quad
|j|\leq\tau\,.
\end{equation}
In particular, the height profile has a macroscopic jump of size
$\tau$ at the boundaries.

\begin{figure}[t!]
\begin{center}
\psfrag{1}[c]{$\tau=1$} \psfrag{2}[c]{$\tau=2$}
\psfrag{3}[c]{$\tau=3$}
\includegraphics[width=\textwidth]{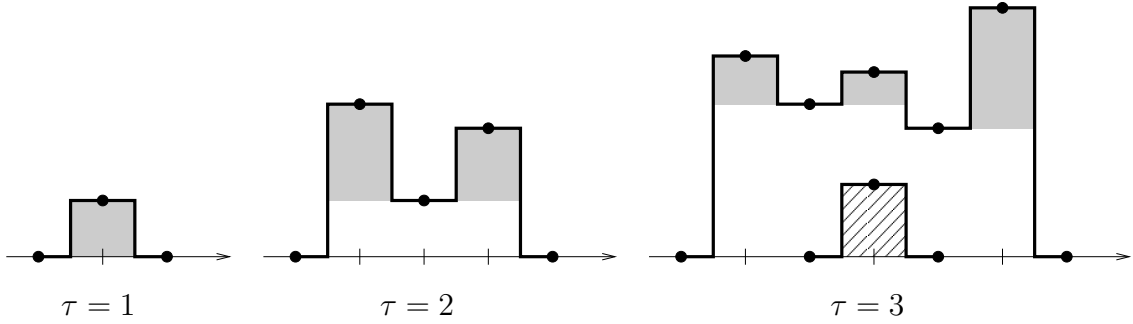}
\caption{A line ensemble for Laguerre growth with shading `grey'
for nucleated mass, `white' for sidewards growth from previous
time step, and `dashed' for excess mass from the line
above.}\label{Fig10}
\end{center}
\end{figure}

As displayed in Figure \ref{Fig10}, the dynamics can be visualized
by extending $\widetilde{h}(j,\tau)$ to a piecewise constant
function with steps on the shifted lattice
$\mathbb{Z}+\frac{1}{2}$. In the random deposition step the
sequence $w(i,\tau+1-i)$, $i=1,\ldots,\tau$, is added at every
second site from left to right  to the current height profile
$\widetilde{h}(j,\tau-1)$. In the deterministic growth, up-steps
move one lattice unit to the left and down-steps to the right.
Thereby neighboring steps overlap and the corresponding excess
mass is deleted.

This last rule is the door for the extra book-keeping heights
$h_\ell(j,\tau)$. Initially $h_\ell(j,0)=0$, $\ell=0,-1,\ldots$.
We set $h_0(j,\tau)=\widetilde{h}(j,\tau)$. Random deposition
takes place only at the top height. The sidewards growth is
carried out simultaneously for all height lines and, at the end of
the sidewards growth, the excess mass at line $\ell$ is copied and
added at the same location to line $\ell-1$. In formulas one sets
\begin{eqnarray}\label{5.8}
h_0(j,\tau)&=&\widetilde{h}(j,\tau),\nonumber \\
h_\ell(j,0)&=&0,\\
h_{\ell-1}(j,\tau+1)&=& \left\{\begin{array}{ll}
h_{\ell-1}(j,\tau)-h_\ell(j,\tau)& \\
+\min\{h_\ell(j-1,\tau), h_\ell(j+1,\tau)\}, & \textrm{if }(-1)^{\tau+j}=1,\\[6pt]
h_{\ell-1}(j,\tau), & \textrm{if }(-1)^{\tau+j}=-1,
\end{array}\right. \nonumber
\end{eqnarray}
for the line labels $\ell=0,-1,\ldots$.

As for the PNG droplet it is possible to describe the statistics  of the collection of points
$\{h_\ell(j,\tau)\,|\,\ell\in\mathbb{Z}_-$, $|j|<\tau$,
$h_\ell(j,\tau)>0\}$ directly without recourse to the stochastic
dynamics as follows.  First we have to define admissible point configurations.
Let $\{x_j, j=-n,\ldots,0\}$ be points on $[0,\infty)$ ordered as
$0\leq x_{-n} \leq \ldots \leq x_0$. We say that $\{x_j,
j=-n,\ldots,0\}\prec\{x_j', j=-n,\ldots,0\}$ if $x_0\leq  x'_0$,
$x_j \leq x'_j \leq  x_{j+1}$ for $j=-n,\ldots,-1$. Admissible point
configurations of the TASEP line ensemble have to satisfy
\begin{eqnarray}\label{5.9}
h_\ell(\pm\tau,\tau)&=&0, \\
\{h_\ell(j,\tau),\ell\in\mathbb{Z}_-\}&\prec&
\{h_\ell(j+1,\tau),\ell\in\mathbb{Z}_-\}, \textrm{ if }|j|<\tau
\textrm{ and }(-1)^{j+\tau}=-1,\nonumber \\
\{h_\ell(j,\tau),\ell\in\mathbb{Z}_-\} &\succ&
\{h_\ell(j+1,\tau),\ell\in\mathbb{Z}_-\}, \textrm{ if }|j|<\tau
\textrm{ and }(-1)^{j+\tau}=1. \nonumber
\end{eqnarray}
As with the growth dynamics, the order $\prec$ and $\succ$ can be
visualized by extending $h_\ell(j,\tau)$ to $\mathbb{R}$ by
setting $h_\ell(x,\tau)=h_\ell(j,\tau)$ for $j-\frac{1}{2}\leq
x<j+\frac{1}{2}$. Then (\ref{5.9}) means that the lines
$h_\ell(x,\tau)$ do not intersect when considered as lines in the
plane, see Figure \ref{Fig10}.

To a given point configuration, \emph{alias} line ensemble, one
associates a weight, which is the product of the weights for each
single step, which in turn equals $e^{-|\delta|/2}$ where $\delta$
is the step size. The total weight is normalized to become a
probability, which then agrees with the probability from the
growth dynamics (\ref{5.8}) at time $\tau$.

To the line ensemble one associates the point process
\begin{equation}\label{5.10}
\eta_{\tau}(y,j)=\sum_{\ell\leq 0}\delta(h_\ell(j,\tau)-y)\,,\quad
y>0\,,
\end{equation}
where $j\in\mathbb{Z}$ refers to time and $y\in\mathbb{R}_+$ to
space. According to our construction, at $y=0$ there are an
infinite number of points. The point process $\eta_{\tau}$ refers
however only to points with a strictly positive $y$ coordinate.
$\eta_{\tau}$ is determinantal with a correlation kernel $R_\tau$,
which is displayed in Proposition 3.3 of \cite{FS}. For $j=0$,
$\tau=2m+1$, the correlation kernel simplifies to
\begin{equation}\label{5.11}
  R_\tau(0,y;0,y')= K_m^{\mathrm{L}}(y,y')\,,
\end{equation}
where $K_m^{\mathrm{L}}$ is the Laguerre kernel of order $m$. In
terms of the standard Laguerre polynominals $L_n$ of order 0, see
\cite{Ab}, it is defined through
\begin{equation}\label{5.12}
K_m^{\mathrm{L}}(y,y')= \sum^{m-1}_{n=0}L_n(y)L_n(y')e^{-y/2}e^{-y'/2}\,.
\end{equation}
In analogy, the line ensemble
$\{h_\ell(j,\tau)\,,\;\ell\in\mathbb{Z}_-\,,\;j\in\mathbb{Z}\}$ is
called the Laguerre line ensemble.

Not to our surprise we have found again a disordered and an
ordered zone separated by the line (\ref{5.7a}) on the macroscopic
scale. Only in this growth model the lower border is trivially the
line $\{y=0\}$.

There are two obvious questions.\medskip\\
(1) Could one choose for $w(i,j)$ a distribution which is different from the
exponential without losing the determinantal property? One can,
but the only admissible modification is $w(i,j)$ to have a
geometric distribution, $\mathbb{P}(w(i,j)=n)=(1-q)q^n$, $0<q<1$,
$n\in\mathbb{N}$, compare with example (iv) of the addendum to
Section \ref{sec.2}. Note that thereby one has constructed a family of growth
models, mostly referred to as discrete time TASEP, which
interpolate between PNG and TASEP. In the limit of rare events,
$q\to 0$, the $w(i,j)$ turn into a Poisson process on
$\mathbb{R}_+^{\,2}$ with constant intensity, while for $q\to 1$, and
proper rescaling, the geometric distribution turns into the
exponential one.\medskip\\
(2) Assuming that $w(i,j)$ is exponential, is it required that
they all have the same mean? In fact not, but the determinantal
property requires a product structure. More concretely, we assume
that $w(i,j)$ are independent exponentials with mean $\langle
w(i,j)\rangle=(a_{ij})^{-1}$. Then it is required that
\begin{equation}\label{5.13}
a_{ij}=a_i+b_j>0\,.
\end{equation}
Before we discussed the special case $a_i=\frac{1}{2}$,
$b_j=\frac{1}{2}$. Our construction of the line ensemble can be
repeated in general, only the weights of the line ensemble have to
be modified. The up-steps are ordered from right to left and the
$j$-th up-step has weight $e^{-a_j|\delta|}$, $\delta$ the step
size, while the down-steps are ordered from left to right with the
$j$-th down-step having weight $e^{-b_j|\delta|}$.


\section{Random matrices and Dyson's Brownian motion}\label{sec.6}
\setcounter{equation}{0}

Random matrices is the most unlikely, and at first sight
unexpected,  item in our list. In retrospect the dynamic exponent
$z=3/2$ of the KPZ equation in one dimension is the ``same'' as
the exponent $1/2$ for the edge of the density of states according
to the Wigner semicircle law. A clear evidence for the link was
established by Johansson \cite{Jo}. He proved that, in the droplet
geometry, the TASEP height above the origin has a scaling function
given by the Tracy-Widom distribution, which was obtained prior by
Tracy and Widom \cite{TW} as the scaling function for the location
of the largest eigenvalue of the Gaussian unitary ensemble (GUE).
Historically, it was a big riddle why the same scaling function
appears in such a diverse context. Our resolution is on the
mathematical side. Largest eigenvalue and height above the origin
result from the edge scaling of a determinantal space-time
process.

The GUE of random matrices is a Gaussian probability distribution
for $N\times N$ complex hermitian matrices defined through
\begin{equation}\label{6.1}
Z^{-1}_N \exp[-\textrm{tr}A^2/2N]\,.
\end{equation}
Here $A=A^\ast$ is a $N\times N$ complex hermitian matrix.
(\ref{6.1}) is understood as a density relative to the flat
measure $dA$ on the independent coefficients of $A$,
\begin{equation}\label{6.1a}
dA=\prod^N_{i=1}d A_{ii}\prod_{1\leq i<j\leq N}
d\mathfrak{R}(A_{ij}) d\mathfrak{I}(A_{ij})\,,
\end{equation}
and $Z_N$ is the normalizing partition function. In convential
random matrix theory the factor $1/2N$ in the exponential is taken
to be 1. Our units are such that the typical spacing between
eigenvalues is of order 1, in accordance with our previous
examples of point processes. Let $\lambda_1,\ldots,\lambda_N$ be
the eigenvalues of $A$. As a consequence of (\ref{6.1}) their
joint probability density is given by
\begin{equation}\label{6.2}
Z^{-1}_N \;|\Delta_N(\lambda)|^2\prod^N_{j=1}e^{-\lambda^2_j/2N}
\end{equation}
with the Vandermonde determinant
\begin{equation}\label{6.3}
\Delta_N(\lambda)=\det((\lambda_i)^{j-1})_{1\leq i,j \leq N}=
\prod_{1\leq i<j\leq N}(\lambda_j-\lambda_i)\,.
\end{equation}
We regard $\lambda_1,\ldots,\lambda_N$ as point process by setting
\begin{equation}\label{6.4}
\eta_N(x)=\sum^N_{j=1}\delta(x-\lambda_j)\,.
\end{equation}
$\eta_N(x)$ is determinantal with correlation kernel given by the
Hermite kernel (\ref{2.22}) with $\widetilde{t} = 2N$.

To make contact with line ensembles one has to advance from the
static distribution (\ref{6.1}) to dynamics. A natural candidate
is the linear Langevin equation,
\begin{equation}\label{6.5}
\frac{d}{dt}A(t)=-\frac{1}{2N}A(t)+\dot{B}(t)\,,
\end{equation}
where $B(t)$ is $N\times N$ complex hermitian matrix-valued
Brownian motion. To say, $B(0)=0$, $\langle B(t)\rangle=0$, and
$B_{ij}(t)$ are complex-valued Gaussian processes with
$B_{ij}(t)^\ast=B_{ji}(t)$ and independent increments,
\begin{equation}\label{6.6}
\langle \dot{B}_{ij}(s)\dot{B}_{i'j'}(t)^\ast\rangle
=\delta(t-s)\delta_{ii'}\delta_{jj'}\,,
\end{equation}
$i,j,i',j'=1,\ldots,N$. Clearly, if in (\ref{6.5}) $A(0)$ is
Gaussian, then so is $A(t)$.

The stationary distribution for (\ref{6.5}) is the GUE probability
measure (\ref{6.1}). Thus a natural choice is to consider the stationary
process for (\ref{6.5}). The eigenvalues
$\lambda_1(t),\ldots,\lambda_N(t)$ of $A(t)$ never intersect and
form a determinantal line ensemble with correlation kernel
\begin{equation}\label{6.7}
  R_N(x,t;x',t')=
  \begin{cases}
    \langle x|e^{-tH_N}P_N e^{t'H_N}|x'\rangle & \text{for }t\geq t'\,, \\
    -\langle x|e^{-tH_N}(1-P_N)e^{t'H_N}|x' \rangle& \text{for }t<t'\,.
  \end{cases}
\end{equation}
Here $H_N$ is the harmonic oscillator Hamiltonian with frequency
$1/2N$,
\begin{equation}\label{6.8}
  H_N=\frac{1}{2}\Big(\! -\frac{d}{dx^2}+\frac{1}
{(2N)^2}\,x^2-\frac{1}{2N}\Big)
\end{equation}
and $P_N$ is the Hermite kernel, i.e., $P_N$ is the projection onto
the first $N$ eigenstates of $H_N$.

As in the previous models, the same line ensemble can be constructed statically. One starts
with $N$ independent Ornstein-Uhlenbeck processes governed by
\begin{equation}\label{6.9}
  \frac{d}{dt}y_j(t)=
  -\frac{1}{2N}y_j(t)+\dot{b}_j(t)\,,
\end{equation}
$j=1,\ldots,N$, with $\{\dot{b}_j(t)\,,\;j=1,\ldots,N\}$ a
collection of $N$ independent white noises. In the time window
$t\in [-\tau,\tau]$ one conditions on the $y_j(t)$'s not to
intersect. The resulting process is denoted by
 $y^{(\tau)}_j(t)$.
Taking the limit
$\tau\to\infty$ one arrives at the by construction stationary diffusion process
$\{y^{(\infty)}_j(t)\,,\;j=1,\ldots,N\,,\; t\in\mathbb{R}\}$. It is
indeed determinantal with correlation kernel (\ref{6.7}).

From the perspective of the PNG droplet, stationarity looks
unnatural. Closer to PNG would be the watermelon ensemble from
Section \ref{sec.2}. In terms of random matrices one sets
\begin{equation}\label{6.10}
A(t)=B(t)-\frac{t}{T}B(T)\,,
\end{equation}
i.e., each matrix element is a Brownian bridge, in particular
$A(0)=0=A(T)$. The eigenvalues $\lambda_1(t),\ldots,\lambda_N(t)$
of $A(t)$ are determinantal with correlation kernel (\ref{2.25}).


\section{Boundary sources}\label{sec.7}
\setcounter{equation}{0}

From the perspective of growth processes the method developed so
far has two drawbacks. Firstly, while we have rather concise
formulas for spatial correlations at fixed growth time $T$, there
is no information on correlations in growth time. This limitation
is intrinsic, since the line ensemble is constructed separately
for each $T$. Secondly, one can allow only for very special
initial conditions which result in surfaces with a nonvanishing
macroscopic curvature. While there is some interest, for example
the Eden growth starting from a single seed builds up an
essentially circular shape, in most computer simulations the
initial condition is a flat surface, which then stays flat on
average.

The restriction to curved profiles can be overcome, at least
partially through the method of boundary sources, which covers
several cases of interest. Boundary sources can be introduced for
PNG, TASEP, and GUE. To avoid repetition we explain only the PNG
model, which happens to be the most transparent case. More details 
are provided in the recent survey \cite{FP05}. The flat
initial height has been resolved only recently \cite{Sa}, see also
\cite{Fe,FS05}. It is tricky with extra ideas and therefore
slightly outside this overview.

We start with the PNG droplet, as explained in Section
\ref{sec.4}, and add additional nucleation events at the two
borders of the sample, i.e., at $x=\pm T$. The sources are Poisson
in time with left rate $\alpha_-$ and right rate $\alpha_+$.
Clearly, the sources will modify the macroscopic shape. But this
is not yet on the agenda. Rather, we want to understand how the
extra sources modify the line ensemble. Switching to the directed
polymer, the sources generate additional nucleation events on the
line $\{v=0\}$ with the intensity $\alpha_+$ and on the line
$\{u=0\}$ with the intensity $\alpha_-$. In the discrete setting,
see Section \ref{sec.5}, the exponential random variables would be
modified such that $\langle w(i,1)\rangle=\alpha_+$, $\langle
w(1,j)\rangle=\alpha_-$, and $\langle w(i,j)\rangle=1$ otherwise.
Note that this modification respects the product form, if the
vectors $\vec{a}$, $\vec{b}$ are altered only in their first entry
from  $\frac{1}{2}$ to $a_1=\alpha^{-1}_+ - \frac{1}{2}$,
$b_1=\alpha^{-1}_- - \frac{1}{2}$. Hence $\langle
w(1,1)\rangle=(\alpha_++\alpha_-
-\alpha_+\alpha_-)/\alpha_+\alpha_-$. Taking the limit of rare
events we conclude that the line ensemble for the PNG model with
boundary sources is still determinantal, provided there is an
extra nucleation event at $(0,0)$ with geometric weight of
parameter $ \alpha_+\alpha_-$.

A further, physically natural choice would be to place a single
source with intensity $\beta$ at $x=0$. In terms of the directed
polymer there are now additional Poisson points along the diagonal
$\{u=v\}$, which should be viewed as a random pinning potential.
For large $\beta$ the directed polymer stays order 1 close to the
diagonal. Any deviation would be too costly energy-wise. As
$\beta$ is decreased there will be longer and longer excursions
away from the diagonal until the critical point
$\beta_{\mathrm{c}}$, when the directed polymer depins. It is
conjectured that $\beta_{\mathrm{c}}=0$ \cite{JL,BSSV}, but there
are counterclaims mostly based on numerical simulations of the
TASEP \cite{Ni}. Unfortunately the source at $x=0$ is not covered
by our methods, since it does not respect the product structure.
To have a determinantal line ensemble one can allow for a general
intensity $\rho(u,v)$ of nucleation events provided it is of the
form $\rho(u,v)=\rho_+(u)\rho_-(v)$, which can be satisfied for
the boundary sources but not for the centered source.

For the PNG model with boundary sources the line ensemble
$\{h_\ell(x,T)$, $\ell=0,-1,\ldots$, $|x|\leq T$\} is constructed
according to the rules of Section \ref{sec.4}. Since the source is
in operation only for $h_0$, one still has $h_\ell(\pm T,T)=\ell$,
$\ell=-1,-2,\ldots$. If $h_0(\pm T,T)=n_{\pm}$, then the
corresponding weight is $(\alpha_+)^{n_+}(\alpha_-)^{n_-}$. This
looks diverging for $\alpha_+,\alpha_-\geq 1$. However the
up-steps and down-steps still carry a $dx$ volume element. Since
they are ordered, one obtains a factor $1/n!$ in the partition
function which makes the total weight finite for all
$\alpha_+,\alpha_-\geq 0$.

We do not provide the details for computing the correlation
kernel, see \cite{FS} for the TASEP. There is however one element
which we want to point out. In the fermion formalism one has a
product of transfer matrices, $e^{-tG}$, and a few number
operators, like $a^\ast(j)a(j)$, sandwiched between the right and
left vectors $\Omega_+,\Omega_-$. If $\alpha_+=\alpha_-=0$, the
boundary conditions are $h_\ell(\pm T,T)=\ell$ which translate to
$\Omega_+=\Omega$, $\Omega_-=\Omega$, where $\Omega$ is the state
with sites $j\leq 0$ occupied and sites $j>0$ empty. If
$\alpha_+>0$, then only the right end point of the top line $h_0$
is lifted upwards. Thus the boundary state becomes
\begin{equation}\label{7.1}
\Omega_+=a^\ast(\psi^+)\widetilde{\Omega}\,,\quad
a^\ast(\psi)=\sum_{j\in\mathbb{Z}}\psi_j
a^\ast(j)\,,\quad\psi^+_{j}=(\alpha_+)^j\,,
\end{equation}
correspondingly for $-$, where $\widetilde{\Omega}$ is the state
with sites $j<0$ occupied and sites $j\geq 0$ empty. For the PNG
model the generator $G$ is the second quantization of nearest
neigbor hopping, which implies that
\begin{equation}\label{7.2}
e^{-tG}a^\ast(\psi^+)e^{tG}= e^{t(\alpha_++
\alpha_+^{-1})}a^\ast(\psi^+)\,.
\end{equation}
Hence the boundary creation operator can be moved from the border to the number
operator $a^\ast(j)a(j)$.

Let us illustrate this simplification by computing the correlation
kernel $R_{\alpha_+,\alpha_-}$ at $t=0$. From Section \ref{sec.4}
we know that for $\alpha_+=0=\alpha_-$
\begin{equation}\label{7.3}
R_{0,0}(j,j')=B_T(j,j')
\end{equation}
with $B_T$ the Bessel kernel. In general one has to compute
expectations of the form
\begin{eqnarray}\label{7.4}
&&Z^{-1}\langle\widetilde{\Omega}|e^{-TG}
a(\psi^-)\prod^m_{k=1}a^\ast(j_k)a(j_k)
a^\ast(\psi^+)e^{-TG}|\widetilde{\Omega}\rangle_\mathcal{F}\,,\nonumber\\
&&Z=\langle\widetilde{\Omega}|e^{-TG}
a(\psi^-)a^\ast(\psi^+)e^{-TG}|\widetilde{\Omega}\rangle_\mathcal{F}\,.
\end{eqnarray}
This results in a determinantal point process with correlation
kernel
\begin{eqnarray}\label{7.6}
&&\hspace{-25pt}R_{\alpha_+,\alpha_-}(j,j')=
B_T(j,j')\nonumber\\&&\hspace{00pt}+\big(\alpha_+\alpha_-\langle\psi^-|
(1-B_T)|\psi^+\rangle\big)^{-1}
((1-B_T)\psi^-)_j((1-B_T)\psi^+)_{j'}\,.
\end{eqnarray}
The boundary sources modify the correlation kernel through a
one-dimensional projection operator. Thus computationally the
resulting difficulties are increased only slightly.

Even without computation one can guess typical configurations of
the line ensemble. To compute $h_0(\pm T,T)$ in terms of the
directed polymer, it has to reach $(2T,0)$, resp.~$(0,2T)$.
Therefore $h_0(\pm T,T)\simeq\alpha_\pm T$. For the line with
label $-1$, just below the top line, we need the extra information
on how $h_{-1}(x,T)$ translates to the directed polymer. It turns
out that for $h_0(x,T)+h_{-1}(x,T)$ one needs to consider two
directed polymers, both starting at $(0,0)$ and ending at
$(x+T,x-T)$. They are required to visit disjoint Poisson points.
Then
\begin{equation}\label{7.5}
h_0(x,T)+h_{-1}(x,T)=
\max_{\begin{subarray}{c} \gamma_1\neq\gamma_2\\ \gamma_1:(0,0)\to(x+T,x-T)\\
\gamma_2:(0,0)\to(x+T,x-T)
\end{subarray}}
\Big(E(\gamma_1)+E(\gamma_2)\Big)\,.
\end{equation}
An according formula holds for $h_0(x,T)+\ldots+h_\ell(x,T)$. If
$x$ is near $\pm T$, the second directed polymer has almost no
Poisson points to visit. Thus $h_{-1}(x,T)\simeq
2T(1-(x/T)^2)^{1/2}$ and the lines with $\ell \leq -1$ form a
disordered zone as before. If $\alpha_+,\alpha_-$ are small, then
$h_0(x,T)$ will follow closely $h_{-1}(x,T)$ in such a way as to
join tangentially the droplet. On the other hand for
$\alpha_+,\alpha_-$ large,
$h_0(x,T)\simeq((\alpha_+-\alpha_-)x+(\alpha_++\alpha_-)T)/2$.
Clearly the most intriguing case occurs when $h_0(x,T)$ is still a
line segment but touches tangentially the droplet. At the touching
point, $x=x_\mathrm{m}$, $h(x_\mathrm{m},T)$ is expected to have
unusual fluctuations. In the picture of the directed polymer, it
chooses either one of the two boundaries and the fluctuations from
the boundary portions are comparable in size to the ones coming
from the bulk. Such fluctuation properties are studied for PNG in
\cite{PS2} and for the TASEP in \cite{FS}.


\section{Edge scaling}\label{sec.8}
\setcounter{equation}{0}

For growth processes the physical height corresponds to the top
line of the line ensemble. Similarly, the facet edge of the Ising
corner is encoded by the top gradient line, see Figure \ref{Fig6}.
Thus our task is to understand the statistical properties of
$h_0$. The most basic information is the size of typical
fluctuations of $h_0$ for large $T$, which defines the
\textit{scaling exponents}, and more precisely the scale invariant
probability distributions for large $T$, which defines the
\textit{scaling functions}. Of course, the hope is that these
quantities do not depend on the details of the line ensemble and
thus are valid for {\it all} line ensembles discussed so far. This
is not so unlikely, since the top line has a lot of space for
fluctuations, which tend to wash out microscopic details. As
guiding example serves a general step random walk, which on a
large scale looks like Brownian motion with the variance of the
step distribution retained as only information on the random walk.
Even more ambitiously we expect that, e.g., in the case of growth
models, the scaling exponents and the scaling functions computed
here are valid for all growth models in the KPZ universality
class. This is in complete analogy to critical phenomena, where
models fall into distinct universality classes. As a rather common
feature, concrete computations can be carried out only for one
specific member of a class.

The finite $T$, resp.~finite $N$, line ensemble is
determinantal. If we consider $T\to \infty$ and focus our
attention on a domain close to the edge, the line statistics there
must be still determinantal. In other words, we only have to study
the limit $T\to \infty$ of the correlation kernel with an
appropriate scaling of its arguments. Through the determinantal
property one deduces the limiting probability distributions from
the limiting correlation kernel.

To illustrate how the scheme works let us consider the PNG model
in the droplet geometry. The starting point is the discrete Bessel
kernel (\ref{4.4}) which is the projection onto all negative
energy states of $\mathfrak{h}_T$ from (\ref{4.3}). For simplicity
let us study the droplet close to $x=0$. Then $\langle
h(0,T)\rangle=2T$ for large $T$ and in the line ensemble we
consider the window $j=2T+yT^\beta$ and $t=\tau T^\alpha$ with
$y,\tau$ of order one and $\alpha$, $\beta$ to be determined.
Inserting in (\ref{4.4}) and switching to the variable $y$ one
arrives at
\begin{equation}\label{8.1}
(\mathfrak{h}_T\psi)(y)=-\psi(y+T^{-\beta})-\psi(y-T^{-\beta})+
\frac{1}{T}(2T+yT^\beta)\psi(y)\,.
\end{equation}
To have a limit one must set
\begin{equation}\label{8.2}
\beta=\frac{1}{3}
\end{equation}
and obtains for $T\to\infty$
\begin{equation}\label{8.3}
T^{2/3}\mathfrak{h}_T\psi(y)=\Big(\!-\frac{d^2}{dy^2}+y\Big)\psi(y)\,.
\end{equation}
Thus under edge scaling $T^{2/3}\mathfrak{h}_T$ goes over to the Airy
operator
\begin{equation}\label{8.4}
H_{\mathrm{Ai}}=-\frac{d^2}{dy^2}+y\,.
\end{equation}
The Airy operator has $\mathbb{R}$ as spectrum with the Airy
function Ai as generalized eigenfunctions
\begin{equation}\label{8.5}
H_\mathrm{Ai}\mathrm{Ai}(y-\lambda)=\lambda\mathrm{Ai}(y-\lambda)\,,
\end{equation}
see \cite{Ab}. In particular the projection onto the eigenstates with negative energies is the
Airy kernel
\begin{equation}\label{8.5a}
 K_\mathrm{Ai}(y,y')=\int^\infty_0 d\lambda
\mathrm{Ai}(y+\lambda)\mathrm{Ai}(y'+\lambda)\,.
\end{equation}
As established with rigor in \cite{PS}, one concludes that
\begin{equation}\label{8.6}
\lim_{T\to\infty} T^{1/3}B_T([2T+yT^{1/3}],[2T+y'T^{1/3}])=
K_\mathrm{Ai}(y,y')
\end{equation}
pointwise.

To have the extended kernel, see (\ref{4.6}), one needs the
scaling limit of $ e^{-t\mathfrak{g}}$ with $t=\tau T^\alpha$.
Since by the argument above the spatial scale is fixed as $T^{1/3}$,
one infers
\begin{equation}\label{8.7}
\alpha=\tfrac{2}{3}\,,\quad \exp[-t\mathfrak{g}]\cong
\exp[\tau(d^2/dy^2)]\,.
\end{equation}
 While the value for
$\alpha$ is correct, the complete asymptotic analysis shows that
the time propagation is governed by the Airy operator,
\begin{eqnarray}\label{8.8}
&&\hspace{-35pt}\lim_{T\to\infty}T^{1/3}B_T\big([2T+T^{1/3}(y-\tau^2)]\,,\;
T^{2/3}\tau\,;\;[2T+T^{1/3}(y'-\tau'^2)]\,,\;T^{2/3}\tau'\big)\nonumber\\
 &&\hspace{0pt}=  \begin{cases}
    \langle y|e^{-\tau H_\mathrm{Ai}}K_\mathrm{Ai}e^{\tau'H_\mathrm{Ai}}|y'\rangle &
    \text{for }\tau\geq\tau'\,, \\
  -\langle y|e^{-\tau H_\mathrm{Ai}}(1-K_\mathrm{Ai})e^{\tau'H_\mathrm{Ai}}|y'\rangle &
  \text{for }\tau<\tau'
  \end{cases}\nonumber\\
&&= K_\mathrm{Ai}(y,\tau;y',\tau')\,.
\end{eqnarray}

The right hand side of (\ref{8.8}) is the extended correlation
kernel of a determinantal process, which we denote by $\xi(y,\tau)$.
$\xi(y,\tau)$ for fixed $\tau$ is concentrated on a discrete set of points,
whose density vanishes as
\begin{equation}\label{8.8a}
\langle\xi(y,\tau)\rangle =
\frac{17}{96\pi}y^{-1/2}\exp[-4y^{3/2}/3]
\end{equation}
for $y\to\infty$ and increases as
\begin{equation}\label{8.8b}
\langle\xi(y,\tau)\rangle \simeq
\frac{1}{\pi}|y|^{1/2}-\frac{1}{4\pi|y|}\cos(4|y|^{3/2}/3)
\end{equation}
for $y\to-\infty$. As a function of $\tau$, $\xi(y,\tau)$ is
concentrated on non-intersecting continuous lines, i.e.,
\begin{equation}\label{8.8c}
\xi(y,\tau)=\sum^0_{j=-\infty}\delta\big(y-y_j(\tau)\big)
\end{equation}
with $\tau\mapsto y_j(\tau)$ continuous. Since
$[H_{\mathrm{Ai}},K_{\mathrm{Ai}}]=0$, $\xi(y,\tau)$ and the
$y_j(\tau)$'s are stochastic processes  stationary in $\tau$.

The convergence in (\ref{8.8}) to the extended kernel carries over
to the convergence of the height $h(x,T)$ of the PNG droplet. One
infers that
\begin{equation}\label{8.9}
\lim_{T\to\infty}T^{-1/3}\big(h(\tau
T^{2/3},T)-2T\big)=y_0(\tau)-\tau^2\,.
\end{equation}
Since Airy functions are all over, $y_0(\tau)$ is baptized as Airy
process \cite{PS} and denoted by $\mathcal{A}(\tau)$. Some of its
properties will be discussed in Section \ref{sec.9}. At the moment
we recall that $\tau$ refers to physical space and $T$ to growth
time. $h(0,T)$ increases linearly and has fluctuations of size
$T^{1/3}$. In the spacial domain of size $\tau T^{2/3}$ the height
statistics is governed by the Airy process plus a systematic
downward bending as $-\tau^2$. Since the propagator on the right
hand side of (\ref{8.8}) and the static kernel are given through
$H_\mathrm{Ai}$, the process $\mathcal{A}(\tau)$ is stationary.
This is physically quite reasonable. In every small region of the
droplet one has the same fluctuation statistics, provided the
local curvature (and possibly linear pieces) are properly
subtracted.

For the Ising corner one also obtains the Airy process for edge
fluctuations, as anticipated. But no simple
short cut as for PNG seems to be available.

It is instructive to repeat the heuristic PNG argument for
stationary Dyson's Brownian motion (\ref{6.5}). The confining
potential  is $V(x) = x^2/2N$, which translates to the potential
$U(x)=(x/2N)^2/2$ on the level of the Hamiltonian $H_N$, see
(\ref{6.8}). Its first $N$ levels are filled up which yields
 the Fermi energy
$E_{\mathrm{F}}=1/2$. The largest eigenvalue of Dyson's Brownian
motion is determined by balancing potential and Fermi energy. Hence
\begin{equation}\label{8.10}
U(\lambda_1)=E_{\mathrm{F}}\,,\quad \mathrm{i.e.,}\;\;
\lambda_1=2N\,,
\end{equation}
which is in agreement with the Wigner semicircle law asserting
the asymptotic density of states as
$\pi^{-1}\big(1-(x/2N)^2\big)^{1/2}$, $|x|\leq 2N$. The
determinantal process close to the edge is governed by the
Hamiltonian (\ref{6.8}) \emph{linearized} at $\lambda_1$, i.e., by
\begin{equation}\label{8.11}
  H_\mathrm{l}=-\frac{1}{2}\frac{d^2}{dx^2}+\frac{1}{2N}x\,.
\end{equation}
Scaling as in (\ref{8.1}), one concludes that $\beta=1/3$. The
time direction has correlations on the scale $N^{2/3}$. By
stationarity of Dyson's Brownian motion, the edge eigenvalues are
thus governed by (\ref{8.8}) in the scaling limit $N\to\infty$.

If instead of the potential $\frac{1}{2}x^2$ we choose some other
potential $V(x)$, the GUE generalizes to
\begin{equation}\label{8.12}
  Z^{-1}_N\exp\big[-N \mathrm{tr}(V(A/N))\big]\,,
\end{equation}
where $V$ is taken as even polynomial with positive leading
coefficient. As before, the distance between eigenvalues is of
order 1. (\ref{6.5}) becomes
\begin{equation}\label{8.13}
\frac{d}{dt}A(t)=-\frac{1}{2}V'(A/N)+\dot{B}(t)
\end{equation}
and (\ref{6.8}) is modified to
\begin{equation}\label{8.14}
H_N=-\frac{1}{2}\frac{d^2}{dx^2}+U_N(x/N)\,.
\end{equation}
$U_N$ has to be chosen such that $V=-\log \psi_\mathrm{g}$, where
$\psi_\mathrm{g}$ is the ground state of $H_N$. $U_N$ depends only
weakly  on $N$. For the construction of the determinantal process
one fills the first $N$ levels of $H_N$, which results in a Fermi
energy $E_\mathrm{F}=\mathcal{O}(1)$. The edge, $x_\mathrm{e}$, is
determined through $U(x_\mathrm{e}/N)=E_\mathrm{F}$. If
$U'(x_\mathrm{e})\neq 0$, then the edge statistics is governed by
the Airy operator. It may happen that $U'(x_\mathrm{e})=0$, but
$U''(x_\mathrm{e})\neq 0$, say. Then the edge statistics changes
and is governed by the Pearcey process, see \cite{TW05} for a
detailed study.


\section{Universal fluctuations}\label{sec.9}
\setcounter{equation}{0}

\begin{figure}[t!]
\begin{center}
\psfrag{b1}[rB]{$\beta=1$}
\psfrag{b2}[rB]{$\beta=2$}
\psfrag{b4}[rB]{$\beta=4$}
\psfrag{s}[c]{$s$}
\psfrag{F}[cB]{$F'_\beta(s)$}
\psfrag{ 0}[cB]{$0$}
\psfrag{ 0.1}[cB]{$0.1$}
\psfrag{ 0.2}[cB]{$0.2$}
\psfrag{ 0.3}[cB]{$0.3$}
\psfrag{ 0.4}[cB]{$0.4$}
\psfrag{ 0.5}[cB]{$0.5$}
\psfrag{ 0.6}[cB]{$0.6$}
\psfrag{-6}[c]{$-6$}
\psfrag{-4}[c]{$-4$}
\psfrag{-2}[c]{$-2$}
\psfrag{0}[l]{$0$}
\psfrag{ 2}[l]{$2$}
\psfrag{ 4}[l]{$4$}
\includegraphics[bb=80 50 550 400,clip,height=5cm]{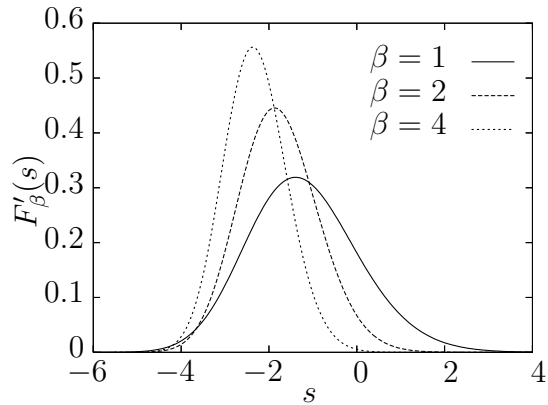}
\caption{Probability densities of the Tracy-Widom
distribution functions at $\beta = 1$ (GOE), $\beta = 2$ (GUE), and
 $\beta = 4$ (GSE). Note that they are not centered.}\label{Fig11}
\end{center}
\end{figure}

Under edge scaling the top line is governed by the Airy process
$\mathcal{A}(\tau)$, which so far was defined only rather
indirectly as $y_0(\tau)$ through (\ref{8.8c}). We return to have a closer look at
its properties. Let us first consider a fixed time, say $\tau=0$.
Then
\begin{equation}\label{9.1}
\mathbb{P}(\mathcal{A}(0)\leq
s)=\mathbb{P}\big(\xi(x,0) \:\textrm{has no point in
}(s,\infty)\big)\,.
\end{equation}
Repeating the computation in (\ref{2.9}), it follows that
\begin{equation}\label{9.2}
\mathbb{P}(\mathcal{A}(0)\leq s)= \det(1-P_sK_\mathrm{Ai}P_s)=
F_\mathrm{GUE}(s)
\end{equation}
with $K_\mathrm{Ai}$ the Airy kernel and $P_s$ the projection onto
$(s,\infty)$. The determinant refers to the Hilbert space
$L^2(\mathbb{R})$. It is well defined, since $P_sK_\mathrm{Ai}P_s$
is of trace class for every $s$. $F_\mathrm{GUE}$ is known as
Tracy-Widom distribution. The corresponding probability
distribution is plotted in Figure \ref{Fig11}. Rather than
computing the determinant one uses that $F_\mathrm{GUE}(s)$ is
related to the Painlev\'{e} II differential equation
\begin{equation}\label{9.3}
u''(s)=2u(s)^3+su(s)\,.
\end{equation}
One picks a special solution, the Hastings-McLeod solution,
uniquely characterized by $u(s)<0$. Then
$F_\mathrm{GUE}(s)=e^{-V(s)}$ with
\begin{equation}\label{9.4}
V(s)=-\int^\infty_s v(x)dx\,,\quad
v(s)=(u(s)^2+s)u(s)^2-u'(s)^2\,.
\end{equation}

During these lectures we have repeatedly raised the issue of the
statistical properties of the layer between ordered and disordered
zones. This can now be answered. The typical width of the border
layer is $T^{1/3}$. Relative to the macroscopic location at one
single point the transverse fluctuations are governed by
$F_{\mathrm{GUE}}$.

The Tracy-Widom distribution was first established in the context
of random matrix theory \cite{TW}. It then appeared in a, at first
sight purely combinatorial problem, namely Ulam's problem, which
poses the question to determine the length of the longest
increasing subsequence of a random permutation \cite{BDJ}, see
also \cite{AD} for a survey. In fact, this problem is identical to
the height $h(0,T)$ of the PNG model. To understand the connection
let us return to the directed polymer, see end of Section
\ref{sec.4}, with starting point $(0,0)$ and end point $(T,T)$.
Let $\omega=(\omega_1,\ldots,\omega_n)$ be a realization of
Poisson points in $[0,T]^2$. We label their 1-coordinates, $x_j$,
in increasing order. Labeling their 2-coordinates, $y_j$, also in
increasing order yields
$\omega=((x_1,y_{\sigma(1)}),\ldots,(x_n,y_{\sigma(n)}))$ and thus
a permutation $(\sigma(1),\ldots,\sigma(n))$ of $(1,\ldots,n)$.
Clearly, $n$ is distributed as $(n!)^{-1}T^{2n} e^{-T^2}$ and for prescribed
$n$ every permutation has the same probability. For given
permutation $\sigma$ we define $\ell_T$ as the length of its
longest increasing subsequence, e.g., the permutation
$(6,2,5,1,4,8,7,3)$ has $\ell_T=3$. From the geometry of the
directed polymer it follows that $\ell_T=e(T,T)$, see (\ref{4.8}),
and hence $\ell_T=h(0,T)$. Thus objects from random matrix theory
made their appearance in growths problems first through the height
above the origin in the PNG model \cite{PS99} and independently
for the TASEP  \cite{Jo}, both in continuous and discrete time. 

More ambitiously, we step to several space points of the PNG droplet 
for large $T$, which means several
fermionic times, $\tau_1<\ldots<\tau_m$, for the Airy process
and consider
\begin{eqnarray}\label{9.4a}
&&\hspace{-15pt}\mathbb{P}(A(\tau_1)\leq
s_1,\ldots,\mathcal{A}(\tau_m)\leq s_m)\nonumber\\
&&
=\mathbb{P}\big(\xi(x_j,\tau_j)\textrm{ has no points in
}(s_j,\infty)\,,j=1,\ldots,m\big)\,.
\end{eqnarray}
Let $K_\mathrm{Ai}$ be the extended Airy kernel of (\ref{8.8}) and
consider
\begin{equation}\label{9.4b}
\theta(y-s_i)K_\mathrm{Ai}(\tau_i,y;\tau_j,y')\theta(y-s_j)
\\=A_{ij}(y,y')
\end{equation}
as a kernel in $L^2(\mathbb{R})\otimes\mathbb{C}^m$. Then, again
by repeating the computation leading to (\ref{2.9}),
\begin{equation}\label{9.5}
\mathbb{P}(\mathcal{A}(\tau_1)\leq
s_1,\ldots,\mathcal{A}(\tau_m)\leq s_m)= \det(1-A)\,.
\end{equation}
Even for $m=2$ the determinant in (\ref{9.5}) cannot be computed
in the simple form as in (\ref{9.3}) and (\ref{9.4}), see however
\cite{AM05,W}, and to extract information requires considerable
effort.

Physically the most robust information is the two-point function
\begin{equation}\label{9.5a}
C(\tau) = \langle \big(\mathcal{A}(\tau) - \mathcal{A}(0)\big)^2\rangle \,.
\end{equation}
For
small $\tau$ Brownian motion dominates and
\begin{equation}\label{9.6}
C(\tau)=2 |\tau|,\quad\textrm{for } \tau\to 0\,.
\end{equation}
 For large
$\tau$ one finds a decay as $|\tau|^{-2}$. In \cite{AM05,W} partial
differential equations for the multi-time distributions of (\ref{9.5}) are
derived, which can be thought of as a generalization of
(\ref{9.3}). As one consequence
\begin{equation}\label{9.7}
C(\tau)=2a_2-2|\tau|^{-2}\,,\quad \tau\gg 1\,,
\end{equation}
with $a_2=0.813...$ the variance of $F_\mathrm{GUE}$.


\section{What have we learned?}\label{sec.10}
\setcounter{equation}{0} {\it One-dimensional growth models in the
KPZ universality class}. We have recovered the dynamical exponent
$z=3/2$, which comes hardly as a surprise, since it is well
established through theoretical arguments and Monte-Carlo
simulations. Novel is the computation of scaling functions along
with the insight that they depend on the geometry of the growth
process \cite{PS00}. If there is a non-zero curvature on the
macroscopic scale, the height fluctuations are governed by the GUE
Tracy-Widom distribution. On the other hand for a macroscopically
flat surface, the scaling function depends on how flat the surface
is prepared initially. The flat surface, no fluctuations at all
initially, has a scaling function different from a surface where
initially the height differences are shortly correlated
\cite{PS,Sa}. It may happen that a flat piece of the surface joins
a curved one. The height fluctuations precisely at the junction
are governed by yet another scaling function. If the surface is
semi-infinite, bordered by a hard wall, the scaling function
changes \cite{IS04}. In this way one realizes the GOE and GSE
random matrix edge scaling distributions, see Figure \ref{Fig11},
{\it and many more}.\medskip\\
{\it Facet edge}. In the scaling limit the fluctuations of the
facet edge are identical to the fluctuations in a growth process
with rounded profile. The linear size of the facet takes here the
role of the growth time $T$. As argued in \cite{FPS04}, the Ising
model with volume constraint and at a temperature below roughening
should have the same facet edge fluctuations. In fact any model
with short range interactions and a non-zero facet edge curvature
is expected to be in the universality class discussed here. There
are other surface models which are still determinantal and exhibit
facets in equilibrium \cite{N87,KOS03}. To establish that their
fluctuation properties are determined by GUE random matrix theory
remains as a task for the future.


\end{document}